\documentclass
[%
reprint,
nofootinbib,
amsmath,amssymb,aps,
prd,
]{revtex4-1}

\usepackage{graphicx}
\usepackage{dcolumn}
\usepackage{bm}
\usepackage{hyperref}
\usepackage{csquotes}
\usepackage{xcolor}
\usepackage{pgfplots}
\usepackage[normalem]{ulem}
\usepackage{mathrsfs}

\begin{document}

\title{Microscopic Cosmological Backreaction of an Ultralight Majoron Coupled to a Coherent Hidden Pseudo-Dirac Ensemble}

\author{Keunsu Cheon}
\affiliation{Graduate School Department of Physics, Daejin University, Pocheon 11159, Korea}

\author{Sin Kyu Kang}
\affiliation{School of Natural Science, Seoul National University of Science and Technology, Seoul 01811, Korea}

\author{Jungjai Lee}
\email{jjlee@daejin.ac.kr}
\affiliation{Graduate School Department of Physics, Daejin University, Pocheon 11159, Korea}
\affiliation{School of Physics, Korea Institute for Advanced Study, Seoul 02455, Korea}

\date{\today}

\begin{abstract}
We develop a microscopic framework for the cosmological backreaction of a hidden pseudo-Dirac fermion ensemble derivatively coupled to an ultralight Majoron. Starting from the effective action, we derive the exact operator Ward identity for the pseudo-Dirac current and the associated microscopic energy-transfer law. For a thermal density matrix prepared at a finite initial time, we calculate the subsequent collisionless, linearized coherence response without assuming continued thermal equilibrium or introducing phenomenological dissipation. The finite-time retarded solution retains the initial boundary transient and nonlocal memory contribution, allowing us to assess the validity of the leading local adiabatic approximation and distinguish reversible coherent energy exchange from dissipative friction. For a subdominant Majoron benchmark evolving in a prescribed Friedmann–Robertson–Walker background, the local energy-transfer rate changes sign, while the present-day fractional velocity correction is only \(+5.23\times10^{-53}\). A convergence-tested numerical comparison, including the initial transient and memory contribution, bounds the difference between the retarded and local fractional velocity corrections to less than \(6.78\times10^{-17}\) of the local correction. Within the linearized, collisionless positive-energy two-state approximation, these results demonstrate that the microscopic backreaction does not produce significant Majoron freezing for the benchmark considered, while leaving open different behavior in other parameter regimes or preparation histories.
\end{abstract}

\maketitle

\section{Introduction}
\label{sec:introduction}

Ultralight scalar fields connect particle physics to cosmological history. The Majoron, denoted by $\phi$, arises from the spontaneous breaking of lepton number and is closely related to Majorana neutrino masses \cite{Chikashige:1981ui,Gelmini:1980re,Schechter:1981cv,Biggio:2026majoronEFT,deGiorgi:2026majoron}.
Small explicit breaking can generate a shallow periodic potential and allow the field to evolve on Hubble time scales \cite{Frieman:1995pm}. Its evolution also depends on its couplings to other particles and on the quantum state of those particles. A microscopic treatment of this dependence is needed before an interaction can be interpreted as slowing or freezing the field. 

Time-dependent Majoron backgrounds have also
been investigated in cosmological leptogenesis
\cite{ChunJung2024,ChunLeeSong:2026}.
Such studies illustrate the cosmological role
of Majoron motion, but do not determine the
backreaction of the prepared hidden
pseudo-Dirac ensemble considered here.

A specific freezing proposal was developed by Simpson
\textit{et al.}\ \cite{Simpson:2018maj}. They showed that persistent energy exchange with neutrinos can prevent a scalar from reversing its motion. In their mechanism, an acceleration-dependent energy transfer can cause rapid slowing, and a derivative neutrino--Majoron
coupling was proposed as a realization. Coherent neutrino motion after the neutrinos become nonrelativistic plays a role in that proposal. These results motivate the question of whether related derivative
couplings can modify scalar evolution. They do not imply that every derivatively coupled fermion ensemble produces freezing. The answer depends on the microscopic current, the quantum state of the fermions, and the resulting energy exchange.

A related proposal explored Majoron freezing induced
by delayed quantum-coherence response in a hidden
pseudo-Dirac fermion sector \cite{Cheon:2026Majoron}.
That study modeled the collective fermionic response
using an effective memory and relaxation framework
and explored the resulting freezing regime.
The present work examines a different dynamical
setting. We specify the initial fermionic density
matrix and calculate the collisionless current response
without imposing a relaxation rate or a phenomenological
energy-transfer law. This allows us to determine
the microscopic backreaction in the stated regime
and to distinguish reversible coherent response
from irreversible friction.

Scalar backreaction is often described through a local friction term or a prescribed energy-transfer rate. Such a description requires a
controlled local limit and a physical mechanism for irreversible energy loss. Linear-response theory distinguishes reactive and
dissipative responses \cite{Kubo:1957}. Microscopic friction can be derived when suitable dynamical and statistical conditions hold,
including in warm-inflation settings \cite{Berera:2008ar}.
Lawrie showed that a time-derivative expansion of nonequilibrium scalar dynamics does not necessarily define a local friction coefficient \cite{Lawrie2003}. A subsequent numerical study showed that
nonequilibrium damping need not be accurately
described by a local friction approximation
\cite{Lawrie:2005friction}. For nearly degenerate fermions, coherence can persist after collisions become negligible. A local damping law therefore cannot be assumed to reproduce their finite-time response.

There is also a distinct, reversible effect. The response of a quantum system can modify the effective inertia of a slowly varying field. Microscopic inertia, geometric response, and nonadiabatic corrections have been studied in general quantum systems
\cite{DAlessioPolkovnikov:2014,KolodrubetzEtAl:2017}.
These studies establish that an inertial correction is not, by itself, evidence for dissipation. They do not determine the response of the initially prepared hidden fermion ensemble considered here.
Moreover, a local expansion need not preserve the boundary transient required by a finite-time initial condition. Both the local reactive response and its deviation from the exact finite-time linear response must therefore be examined.

A pseudo-Dirac fermion consists of two nearly degenerate Majorana mass eigenstates. Their small mass splitting breaks an approximate Dirac-number symmetry. Dirac--Majorana neutrino constraints,
pseudo-Dirac mass structures, and their oscillation phenomenology have been studied previously
\cite{Enqvist:1992dm,Chang:2000pd,Kobayashi:2000md,Biswas:2025pseudoDirac,Carloni:2026quasiDirac}.
Ultralight scalar dark matter can also induce small Majorana masses and pseudo-Dirac behavior in neutrino models \cite{Dev:2023ul}.
In this work the fermions instead form a hidden pseudo-Dirac sector. They are not identified with active neutrinos, and their quantum state is specified independently. This distinction matters because
the current response depends on the prepared occupation difference and on the coherence that develops between the Majorana mass eigenstates.

We study a homogeneous Majoron derivatively coupled to the
pseudo-Dirac vector current. In the Majorana mass basis the current is off diagonal, so its expectation value probes inter-state coherence. An exact operator Ward identity relates the current divergence to the Majorana mass splitting $\delta m\equiv m_1-m_2>0$, where $m_1$ and $m_2$ are the two Majorana masses. The derivative interaction preserves the Dirac-number symmetry; the explicit breaking comes from the Majorana masses. Consequently, the current is conserved in the exact Dirac limit $\delta m\to0$. For nonzero splitting, the Ward identity fixes the operator form of the energy-transfer source, but not its expectation value or sign. Those must be determined from the evolving fermion density matrix. The microscopic energy-transfer law is not a postulated positive-definite friction term; its derivation is given in Sec.~\ref{subsec:scalar_energy_exchange}. 
The quantum state is prepared at a finite cosmic time $t_i$.
We choose a thermal density matrix that is diagonal in the Majorana mass basis, with vanishing initial chemical potentials and inter-state coherence. The Majoron starts at rest, $\dot\phi(t_i)=0$, where a dot denotes a derivative with respect to cosmic time. The derivative drive then vanishes at preparation,
and the initial density matrix commutes with the free mass
Hamiltonian. As the Majoron subsequently moves, the interaction creates coherence dynamically. Asymmetric or two-sector reheating, followed by hidden-sector interactions, can motivate such a prepared
ensemble \cite{AdsheadCuiShelton2016,AdsheadRalegankarShelton2019}.
Those interactions would need to establish the assumed thermal state before becoming negligible. We do not model this earlier preparation process. For $t\geq t_i$ we instead assume collisionless
evolution, without continued thermalization or an imposed dephasing rate. At zeroth order, occupations remain fixed at comoving momentum $q=ap$, where $a$ is the Friedmann--Robertson--Walker (FRW) scale factor and $p$ is physical momentum. For massive fermions, these
fixed occupations are not generally equal to instantaneous thermal distributions at a common redshifting temperature.

Our analysis connects the microscopic current to cosmological backreaction in three steps. First, we construct the normalized current and scalar-bilinear matrix elements in the positive-energy
two-state subspace. We verify their fixed-time Ward relation and identify the terms arising from the time dependence of the spinor basis. This distinguishes the exact operator Ward identity
from the fixed-time Ward relation in the
positive-energy two-state subspace.
The latter does not by itself establish the
full time-dependent Ward identity, because
the spinor basis evolves in an expanding universe. 

Second, we solve the collisionless linearized coherence equation from the specified preparation time.
Its finite-time retarded solution retains both the initial boundary transient and the nonlocal memory integral. Although the derivative drive vanishes at preparation,
its time derivative need not vanish because the
initial Majoron acceleration can be nonzero.
The finite-time boundary transient is then needed
to preserve the prepared initial condition in the
local adiabatic expansion.
Without collisions, the coherent oscillations are not assigned an ad hoc damping rate. A local adiabatic response follows only under additional source-variation and level-splitting conditions, which are distinct from the conditions controlling positive- and
negative-energy mixing. 

Third, we use the induced current to compute the leading Majoron backreaction on a prescribed FRW background.
The local limit yields a reactive inertia correction and an energy- transfer rate whose sign can change. Neither feature by itself establishes irreversible friction.

For our specified subdominant Majoron benchmark,
the calculated local backreaction is extremely small
and does not produce significant freezing. We derive a modewise bound on the difference
between the finite-time retarded coherence
and its leading local adiabatic approximation.
We then integrate this bound over momentum
and propagate the resulting current-difference
bound to the Majoron velocity response using
a nonnegative comparison system. The comparison includes the initial boundary transient and the remaining memory contribution. For the benchmark, the numerical comparison gives a fractional difference between the retarded and local velocity responses below approximately $6.78\times10^{-17}$ of the local
response; see Eq.~\eqref{eq:V_retarded_velocity_bound} and
Appendix~\ref{app:nonlocal_comparison}. Thus, within the stated linearized collisionless calculation, the nonlocal correction does not change the conclusion that significant freezing is absent for this benchmark. The bound is obtained from the comparison system,
not from a direct phase-resolved integration of every rapidly oscillating mode. It does not exclude different outcomes for other parameters or preparation histories.

The calculation has a defined scope. The Ward identity is an exact operator relation within the effective theory. The finite-time retarded solution is exact for the linearized, positive-energy two-state initial-value problem on the prescribed FRW background,
and the scalar backreaction is treated perturbatively. The full coupled nonlinear Einstein--Majoron--fermion evolution and a complete ultraviolet construction are not computed. The Majoron is not assumed to account for the observed dark-energy density.
Further limitations and possible extensions are discussed in
Sec.~\ref{sec:discussion_conclusions}.

The remainder of the paper is organized as follows.
Section~\ref{sec:framework} introduces the effective theory, the operator Ward identity, the energy-transfer law, and the prepared
state. Section~\ref{sec:ward_current} derives the spinor projections and their Ward consistency conditions.
Section~\ref{sec:coherent_kinetics} develops the finite-time retarded response and its local adiabatic limit.
Section~\ref{sec:cosmological_backreaction} evaluates the FRW backreaction and the numerical comparison.
Section~\ref{sec:discussion_conclusions} presents the discussion and conclusions. Appendix~\ref{app:scalar_potential_majoron}
derives the physical Majoron direction
and its periodic potential. Appendix~\ref{app:hidden_pseudo_dirac_origin}
describes the pseudo-Dirac mass structure.
Appendix~\ref{app:four_component_ward} verifies the Ward identity in the complete spinor basis. Appendix~\ref{app:derivative_portal_origin} illustrates a gauge-invariant origin of the derivative portal. Appendix~\ref{app:nonlocal_comparison} gives the finite-time nonlocal comparison, its error bounds, and the numerical convergence tests.

\section{Effective Majoron--Pseudo-Dirac Framework}
\label{sec:framework}

\subsection{Physical Majoron and its potential}
\label{subsec:majoron_sector}

We consider two complex scalar fields $\Phi_1$ and $\Phi_2$
with charges $Q_1=2$ and $Q_2$, respectively, under a gauged
$U(1)_{B-L}$. We take $Q_2\geq 3$ to be an odd integer.
Since $\gcd(2,Q_2)=1$, the lowest-dimensional
gauge-invariant, phase-dependent operator
constructed solely from $\Phi_1$, $\Phi_2$,
and their conjugates has dimension $Q_2+2\geq 5$.
Thus, no renormalizable operator in this
two-scalar sector generates a potential
for the relative phase.

After symmetry breaking, we write
\begin{equation}
\Phi_j(x)=
\frac{v_j+\rho_j(x)}{\sqrt{2}}e^{i\theta_j(x)},
\qquad j=1,2.
\label{eq:Phi_phases}
\end{equation}
Here $v_j$ are the vacuum expectation values,
$\rho_j$ are the radial fluctuations, and $\theta_j$
are dimensionless phases.

One phase combination is absorbed by the
$U(1)_{B-L}$ gauge boson. After removing this
mode and integrating out the heavy radial fields,
the remaining physical Majoron is
\begin{equation}
\vartheta=Q_2\theta_1-2\theta_2,
\qquad
\phi=f_\phi\vartheta,
\qquad
f_\phi=
\frac{v_1v_2}
{\sqrt{4v_1^2+Q_2^2v_2^2}}.
\label{eq:majoron_period}
\end{equation}
Here $\vartheta$ is the gauge-invariant phase and
$f_\phi$ is the Majoron periodicity scale, fixed
by the canonically normalized scalar kinetic term.
Because the charges are coprime, the fundamental
period is $2\pi f_\phi$.

Explicit breaking of the accidental global phase
symmetry generates the periodic Majoron potential
\cite{Babu1993,GhoshLohoManna2026}
\begin{equation}
V(\phi)=\Lambda_\phi^4
\left[1-\cos\left(\frac{\phi}{f_\phi}\right)\right],
\qquad
m_\phi^2=\frac{\Lambda_\phi^4}{f_\phi^2}.
\label{eq:majoron_potential}
\end{equation}
Here $\Lambda_\phi^4$ is the potential amplitude,
and $m_\phi$ is the Majoron mass near a potential
minimum. The leading phase-breaking operator
and the derivation of $\Lambda_\phi^4$ are given
in Appendix~\ref{app:scalar_potential_majoron}.

For the benchmark considered below, we choose
$Q_2=15$, $|\kappa|=1$,
$v_1=1.13\times10^{11}\mathrm{GeV}$, and
$v_2=2.02\times10^9\mathrm{GeV}$. 
These parameters approximately reproduce
$f_\phi=10^9\mathrm{GeV}$ and
$m_\phi=10^{-33}\mathrm{eV}$,
using the non-reduced Planck mass $M_{\rm Pl}$ defined in
Appendix~\ref{app:scalar_potential_majoron}.
The leading phase-breaking operator has dimension 17.
The larger scale $v_1$ can also set the mass scale
of a conventional Type-I seesaw sector.
For right-handed neutrinos with $B-L=-1$,
the Majorana Yukawa interaction
$(y_M/2)\Phi_1 N_R N_R+\mathrm{H.c.}$
gives $M_R=y_Mv_1/\sqrt{2}$.
For $y_M\sim\mathcal{O}(1)$, this yields
$M_R\sim10^{11}\,\mathrm{GeV}$.
Light-neutrino masses of order
$m_\nu\sim0.05\,\mathrm{eV}$ can then be obtained
with a moderate Dirac Yukawa coupling
$y_D\sim10^{-2}$. 

\subsection{Hidden pseudo-Dirac fermions}
\label{subsec:hidden_fermions}

The hidden sector contains two left-handed Weyl fields,
$N$ and $S$. Both are singlets under the gauged
$U(1)_{B-L}$ and are distinct from the right-handed
neutrinos of the usual seesaw sector. Their kinetic
and mass terms are
\begin{align}
\mathcal L_{\rm PD}={}&
iN^\dagger\bar\sigma^\mu\nabla_\mu N
+iS^\dagger\bar\sigma^\mu\nabla_\mu S\nonumber\\
&-\left[
m_DNS+\frac{\mu}{2}(NN+SS)+\mathrm{H.c.}
\right].
\label{eq:hidden_lagrangian}
\end{align}
Here $\bar\sigma^\mu$ are the Weyl spin matrices,
$\nabla_\mu$ is the spinor covariant derivative,
$m_D$ is the Dirac mass, $\mu$ is the common
Majorana mass, and H.c. denotes the Hermitian
conjugate of the mass terms. Spinor contractions
in $NS$, $NN$, and $SS$ are implicit.
We choose real mass parameters with $m_D>\mu>0$.
We take equal Majorana mass parameters for
$N$ and $S$ as a simplifying choice.
A more general mass structure is discussed
in Appendix~\ref{app:hidden_pseudo_dirac_origin}.

We define the two Majorana mass eigenstates by
\begin{equation}
N_1=\frac{N+S}{\sqrt{2}},
\qquad
N_2=\frac{i(N-S)}{\sqrt{2}},
\label{eq:hidden_mass_basis}
\end{equation}
where $N_1$ and $N_2$ are the left-handed components
of the two Majorana mass eigenstates. The factor
of $i$ in $N_2$ makes both Majorana masses positive:
\begin{equation}
\begin{aligned}
m_1&=m_D+\mu,
\qquad m_2=m_D-\mu,\\
\delta m&\equiv m_1-m_2=2\mu>0.
\end{aligned}
\label{eq:hidden_mass_splitting}
\end{equation}
The small splitting is technically natural
\cite{CentellesChulia2021}.
When $\mu\to0$, the hidden Dirac-number symmetry
$U(1)_{\rm PD}$ is restored. Under this symmetry,
$N\to e^{i\beta}N$ and $S\to e^{-i\beta}S$,
where $\beta$ is a constant phase.

The corresponding four-component Majorana fields
are $\Psi_\alpha=(N_\alpha,N_\alpha^\dagger)^T$,
where $\alpha=1,2$ labels the mass eigenstates
and $T$ denotes transpose. The four-component
Dirac field is
\begin{equation}
N_D=
\begin{pmatrix}
N\\
S^\dagger
\end{pmatrix}
=
\frac{\Psi_1-i\Psi_2}{\sqrt{2}},
\label{eq:ND_definition}
\end{equation}
where $N_D$ is the Dirac spinor formed from $N$ and $S$.

\subsection{Derivative portal and effective action}
\label{subsec:portal_current}

We take the leading effective Majoron--hidden-fermion interaction to be
\begin{equation}
\mathcal L_{\phi\rm PD}=\frac{\partial_\mu\phi}{f_{\rm eff}}
J_{\rm PD}^\mu,\qquad
J_{\rm PD}^\mu=:\bar N_D\gamma^\mu N_D:.
\label{eq:derivative_portal}
\end{equation}
This derivative operator is invariant under a constant Majoron shift;
the potential in Eq.~\eqref{eq:majoron_potential} supplies explicit shift symmetry breaking. Since $N$ and $S$ are $B-L$ singlets, the portal is gauge invariant but is not fixed by their $B-L$ charges. In particular,
$f_{\rm eff}$ must not be identified with $f_\phi$ or with a high-scale seesaw vacuum expectation value. 

We treat $1/f_{\rm eff}$ as an effective
Wilson coefficient and restrict the calculation
to energies well below the relevant ultraviolet
completion scale.
Derivative interactions of light pseudoscalar
fields can be described systematically within
effective field theory \cite{Brivio2017,Bauer2021}.
Clockwork constructions provide possible
mechanisms for hierarchies among effective coupling scales
\cite{KaplanRattazzi2016,GiudiceMcCullough2017},
although no specific clockwork realization
is assumed in the present calculation.
The dimension-six construction in
Appendix~\ref{app:derivative_portal_origin}
provides an illustrative gauge-invariant
origin of the derivative portal.
The response calculation below uses the
low-energy coefficient $1/f_{\rm eff}$
as an independent input.

We use the metric signature $(-,+,+,+)$ and a spatially flat FRW background,
\begin{equation}
ds^2=-dt^2+a^2(t)d\mathbf{x}^2,\qquad H=\frac{\dot a}{a},
\label{eq:FRW_metric}
\end{equation}
where $a(t)$ is the cosmological scale factor.
Collecting the scalar and hidden-fermion sectors, the low-energy matter
action is
\begin{equation}
\begin{split}
S_{\rm eff}=\int d^4x\,\sqrt{-g}\,\bigg[
&-\frac12 g^{\mu\nu}\partial_\mu\phi\,\partial_\nu\phi-V(\phi)\\
&+\mathcal L_{\rm PD}
+\frac{\partial_\mu\phi}{f_{\rm eff}}J^\mu_{\rm PD}\bigg].
\end{split}
\label{eq:effective_action}
\end{equation}
Here $\mathcal L_{\rm PD}$ is given in
Eq.~\eqref{eq:hidden_lagrangian}, and the spinor covariant derivatives include the spin connection. The metric is prescribed in the spectator
calculation; Eq.~\eqref{eq:effective_action} is the matter action on that background, not an action from which we solve for the expansion history.

\subsection{Current operator and Ward identity}
\label{subsec:ward_operator}

Using the Dirac field in
Eq.~\eqref{eq:ND_definition}, we express the
current in Eq.~\eqref{eq:derivative_portal}
in the Majorana mass basis.
For Majorana fields, the diagonal vector
bilinears vanish,
$\bar\Psi_\alpha\gamma^\mu\Psi_\alpha=0$.
The off-diagonal bilinears satisfy
$\bar\Psi_2\gamma^\mu\Psi_1
=-\bar\Psi_1\gamma^\mu\Psi_2$.
The current therefore becomes
\begin{equation}
J_{\rm PD}^\mu
=-i:\bar\Psi_1\gamma^\mu\Psi_2:.
\label{eq:III_J_majorana}
\end{equation}
The colons denote normal ordering.
The factor of $i$ follows from the
phase convention in
Eq.~\eqref{eq:hidden_mass_basis}.
It is not an independent normalization
factor.

The same charge density follows directly
from the original Weyl fields:
\begin{equation}
\begin{aligned}
J_{\rm PD}^0
&=:N^\dagger N-S^\dagger S:\\
&=-i:N_1^\dagger N_2:
+i:N_2^\dagger N_1:.
\end{aligned}
\label{eq:III_J_weyl}
\end{equation}
Here the dagger denotes Hermitian conjugation.
This provides an independent check of
the sign and normalization of
Eq.~\eqref{eq:III_J_majorana}.

The divergence of the current is fixed
by the Majorana mass terms, which break
the approximate Dirac-number symmetry.
To identify this contribution, first
set the derivative portal to zero.
For constant real masses, the Majorana equations
and their adjoints give\footnote{
We use $\{\gamma^{\hat a},\gamma^{\hat b}\}
=-2\eta^{\hat a\hat b}$, where
$\eta^{\hat a\hat b}=\mathrm{diag}(-,+,+,+)$
and hatted indices denote a local orthonormal frame.
In the Weyl basis,
$\gamma^{\hat 0}=
\left(\begin{smallmatrix}
0&I_2\\ I_2&0
\end{smallmatrix}\right)$
and
$\gamma^{\hat i}=
\left(\begin{smallmatrix}
0&\sigma^i\\-\sigma^i&0
\end{smallmatrix}\right)$,
where $I_2$ is the $2\times2$ identity matrix
and $\sigma^i$ are Pauli matrices.
We define $\bar\Psi=\Psi^\dagger\gamma^{\hat 0}$
and use
$(i\gamma^\mu\nabla_\mu-m_\alpha)\Psi_\alpha=0$.
These conventions fix the signs of the
Majorana equations and their adjoints.
\begin{equation}
\begin{aligned}
(\nabla_\mu\bar\Psi_1)\gamma^\mu
&=im_1\bar\Psi_1,\\
\gamma^\mu\nabla_\mu\Psi_2
&=-im_2\Psi_2.
\end{aligned}
\label{eq:III_adjoint_eom}
\end{equation}
}
Taking the covariant divergence of
Eq.~\eqref{eq:III_J_majorana} and using
Eq.~\eqref{eq:III_adjoint_eom}, we obtain
\begin{align}
\nabla_\mu J_{\rm PD}^\mu
={}&-i:
\left[
(\nabla_\mu\bar\Psi_1)\gamma^\mu\Psi_2
+\bar\Psi_1\gamma^\mu\nabla_\mu\Psi_2
\right]:
\nonumber\\
={}&(-i)(im_1-im_2)
:\bar\Psi_1\Psi_2:
\nonumber\\
={}&\delta m:\bar\Psi_1\Psi_2:.
\label{eq:III_ward_operator}
\end{align}

This identity also holds when the
derivative portal is included.
Under the Dirac-number transformation
$N_D\to e^{i\beta}N_D$, the current
$J_{\rm PD}^\mu$ is invariant.
Therefore, the derivative interaction
in Eq.~\eqref{eq:derivative_portal}
does not break this symmetry.
Its contributions to the current
divergence cancel.
Only the explicit symmetry-breaking
mass terms contribute to
Eq.~\eqref{eq:III_ward_operator}.
The identity is exact at the operator
level, provided the composite operators
are defined in a symmetry-preserving
renormalization scheme.

For a homogeneous and isotropic state
in the spatially flat FRW background,
$\langle J_{\rm PD}^i\rangle=0$.
The covariant Ward identity then gives
\begin{equation}
\frac{d}{dt}\langle J_{\rm PD}^0\rangle
+3H\langle J_{\rm PD}^0\rangle
=
\delta m
\langle:\bar\Psi_1\Psi_2:\rangle.
\label{eq:III_ward_FRW}
\end{equation}
Here $\langle\cdots\rangle$ denotes
the expectation value in the fermionic
state, and $i=1,2,3$ labels spatial
components.

The term $3H\langle J_{\rm PD}^0\rangle$
accounts for the dilution of the
physical charge density by cosmic
expansion. It is required by the
covariant divergence and is independent
of the preparation of the fermionic
state. It does not represent
dissipative friction or additional
Dirac-number violation.

\subsection{Majoron equation and microscopic energy exchange}
\label{subsec:scalar_energy_exchange}

Varying the effective action in
Eq.~\eqref{eq:effective_action} with respect
to $\phi$ and integrating by parts gives
\begin{equation}
\Box\phi-V_{,\phi}
-\frac{1}{f_{\rm eff}}\nabla_\mu J^\mu_{\rm PD}=0,
\qquad
\Box\equiv\nabla_\mu\nabla^\mu.
\label{eq:phi_eom_covariant}
\end{equation}
Here $V_{,\phi}\equiv dV/d\phi$ is the derivative
of the Majoron potential.

We treat the homogeneous Majoron as a classical
field and evaluate the current in the fermionic
quantum state. Using
$\Box\phi=-\ddot\phi-3H\dot\phi$ and the FRW
Ward identity in Eq.~\eqref{eq:III_ward_FRW}, we obtain
\begin{equation}
\ddot\phi+3H\dot\phi+V_{,\phi}
=
-\frac{1}{f_{\rm eff}}
\left(
\frac{d}{dt}\langle J_{\rm PD}^{0}\rangle
+3H\langle J_{\rm PD}^{0}\rangle
\right).
\label{eq:phi_eom_framework}
\end{equation}
Here $\langle J_{\rm PD}^{0}\rangle$ is the
expectation value of the physical charge
density in the evolving fermionic state.

We define the Majoron energy density and
pressure as
$\rho_\phi=\dot\phi^2/2+V(\phi)$ and
$P_\phi=\dot\phi^2/2-V(\phi)$.
Multiplying
Eq.~\eqref{eq:phi_eom_framework}
by $\dot\phi$ gives
\begin{equation}
\dot\rho_\phi
+3H(\rho_\phi+P_\phi)
=-Q_{\rm micro}.
\label{eq:phi_continuity_framework}
\end{equation}
Here $Q_{\rm micro}$ is the microscopic
energy-transfer rate per unit physical
volume. It is given by
\begin{equation}
\begin{aligned}
Q_{\rm micro}
&=
\frac{\dot\phi}{f_{\rm eff}}
\left(
\frac{d}{dt}\langle J_{\rm PD}^{0}\rangle
+3H\langle J_{\rm PD}^{0}\rangle
\right)\\
&=
\frac{\delta m\,\dot\phi}{f_{\rm eff}}
\left\langle:\bar\Psi_1\Psi_2:\right\rangle.
\end{aligned}
\label{eq:Q_ward_framework}
\end{equation}
The second equality follows
from the exact Ward identity in
Eq.~\eqref{eq:III_ward_FRW}.

We use the convention that $Q_{\rm micro}>0$
denotes energy transferred from the Majoron
to the hidden pseudo-Dirac sector.
A negative value denotes transfer in the
opposite direction.

The action and Ward identity fix the form of
$Q_{\rm micro}$ but not its sign or magnitude.
These depend on the expectation value of the
scalar bilinear in the evolving fermionic state.
The resulting energy transfer is not an assumed
positive-definite friction law.

\subsection{Comoving momentum and the prepared ensemble}
\label{subsec:ensemble_scope}

We label each fermion mode by its conserved comoving
momentum $q=ap$, where $p=q/a$ is the physical momentum.
The mode energies and their splitting are
\begin{equation}
\begin{aligned}
E_{\alpha,q}(t)
&=\sqrt{\frac{q^2}{a^2(t)}+m_\alpha^2},\\
\Delta E_q(t)
&\equiv E_{1,q}(t)-E_{2,q}(t)>0.
\end{aligned}
\label{eq:energies_framework}
\end{equation}
We use $h=\pm1$ to label helicity.
The initial occupations do not depend on $h$,
but helicities are summed explicitly when computing
physical densities.

At a finite preparation time $t_i$, we take
the reduced one-particle density matrix to be
isotropic and diagonal in the Majorana mass basis:
\begin{align}
\rho_{qh}(t_i)
&=
\begin{pmatrix}
f_{1,q}(t_i)&0\\
0&f_{2,q}(t_i)
\end{pmatrix},
\qquad
\rho_{12,qh}(t_i)=0,
\label{eq:initial_rho_framework}\\
f_{\alpha,q}(t_i)
&=
\frac{1}
{\exp[E_{\alpha,q}(t_i)/T_{\rm PD,i}]+1}.
\label{eq:initial_FD_framework}
\end{align}
Here $\rho_{qh}$ is the two-state density matrix,
$f_{\alpha,q}(t_i)$ are its diagonal occupations,
and $T_{\rm PD,i}$ is the hidden pseudo-Dirac
temperature at preparation.

We impose vanishing initial chemical potentials as part of the specified thermal preparation. This choice corresponds to a charge-symmetric initial ensemble and is compatible with thermal and chemical equilibrium when the relevant number-changing processes are effective. However, because the pseudo-Dirac-number symmetry is only weakly broken, vanishing chemical potentials do not follow solely from the Majorana character of the mass eigenstates
\cite{GopalakrishnaTibrewala2025}.
Zero-chemical-potential Fermi--Dirac distributions
are also used in cosmological neutrino kinetic studies
\cite{AkitaYamaguchi2020,FrousteyPitrouVolpe2020}.
We do not model these processes and impose this distribution only at $t_i$.

The prepared occupation difference is fixed
by the mass splitting:
\begin{equation}
\begin{aligned}
D_q(t_i)
&\equiv f_{2,q}(t_i)-f_{1,q}(t_i)\\
&=
\frac{\Delta E_q(t_i)}{T_{\rm PD,i}}
n_F\left(\frac{E_q(t_i)}{T_{\rm PD,i}}\right)\\
&\quad\times
\left[
1-n_F\left(\frac{E_q(t_i)}{T_{\rm PD,i}}\right)
\right]\\
&\quad+
\mathcal O\left[
\left(\frac{\Delta E_q(t_i)}{T_{\rm PD,i}}\right)^3
\right].
\end{aligned}
\label{eq:occupation}
\end{equation}
Here $E_q(t_i)=[E_{1,q}(t_i)+E_{2,q}(t_i)]/2$
is the mean energy at preparation, and
$n_F(x)=(e^x+1)^{-1}$ is the Fermi--Dirac
function of the dimensionless argument $x$.
The expansion requires
$\Delta E_q(t_i)\ll T_{\rm PD,i}$.
Since $m_1>m_2$, the exact difference satisfies
$D_q(t_i)>0$.
It is not an independently chosen initial asymmetry.

The Majoron is initially at rest,
$\dot\phi(t_i)=0$.
The homogeneous derivative drive therefore
vanishes at $t_i$.
The positive-energy two-state Hamiltonian $H_{qh}^{(+)}$,
defined in Eq.~\eqref{eq:III_Hpositive},
is therefore diagonal in the Majorana mass basis
at $t_i$. Since the initial density matrix
is diagonal in the same basis,
$[H_{qh}^{(+)}(t_i),\rho_{qh}(t_i)]=0$.
The collisionless two-state equation then gives
$\dot\rho_{12,qh}(t_i)=0$.
As the Majoron begins to move, the derivative interaction can generate off-diagonal coherence.

The zero initial coherence is defined
in the Majorana mass basis.
A rotation to the $(N,S)$ two-state basis
gives an off-diagonal entry $-D_q(t_i)/2$
in the rotated occupation matrix.
This is a basis effect, not newly produced
Majoron coherence.
Its relation to Weyl-field correlators
requires the spinor projections developed in
Sec.~\ref{subsec:ward_spinor_projection}.

For $t\geq t_i$, we neglect the hidden-sector
collision operator and work to first order
in the Majoron-induced coherence
\cite{KainulainenParkkinen2024}.
Changes in the diagonal occupations start
at second order.
At fixed $q$,
\begin{equation}
f_{\alpha,q}^{(0)}(t)=f_{\alpha,q}(t_i),
\qquad
D_q^{(0)}(t)=D_q(t_i),
\qquad t\geq t_i.
\label{eq:free_streaming_framework}
\end{equation}
Here the superscript $(0)$ denotes zeroth order
in the derivative drive.
These occupations remain fixed at comoving
momentum. For massive fermions, they do not
generally equal instantaneous thermal distributions
at a common redshifting temperature.

A possible hidden mediator $\chi$ could establish
the prepared thermal state before $t_i$.
Here $\chi$ denotes a hypothetical hidden-sector
particle, not an active mediator in the subsequent
evolution. Preparation would require kinetic
and chemical equilibration rates
$\Gamma_{\rm kin}$ and $\Gamma_{\rm chem}$
to satisfy
$\Gamma_{\rm kin},\Gamma_{\rm chem}\gg H$
before decoupling.
For later collisionless evolution, a sufficient
condition is
\begin{equation*}
\int_{t_i}^{t}dt'\,
\Gamma_{\rm coll}(t')\ll1,
\end{equation*}
where $\Gamma_{\rm coll}$ is the relevant
total collision rate.
An ongoing $\chi$ bath with appreciable
scattering or pair production would require
explicit collision terms.
We assume an idealized decoupling and do not
calculate its rates or introduce a
phenomenological dephasing term.

For later use, we record the positive-energy
representation of the pseudo-Dirac charge density.
Let $\mathsf Q$ be its unprojected one-particle
matrix and let $U_{qh}$ have the two normalized
positive-energy Majorana spinors as columns.
Then
\begin{align}
U_{qh}^\dagger\mathsf Q U_{qh}
&=A_q\sigma_y,
\label{eq:A_projection_framework}\\
A_q
&=\left[
\frac{E_{1,q}E_{2,q}+p^2+m_1m_2}
{2E_{1,q}E_{2,q}}
\right]^{1/2}.
\label{eq:A_exact_framework}
\end{align}
Here $\sigma_y$ is the second Pauli matrix
in the Majorana two-state space, and $A_q$
is the normalized overlap of the two
positive-energy spinors.
It satisfies $A_q\to1$ as $\delta m\to0$.
The derivation and its Ward-identity check
are given in
Sec.~\ref{subsec:ward_spinor_projection}.

\subsection{Benchmark scales and domain of validity}
\label{subsec:benchmark_scope}

For the hidden pseudo-Dirac fermions and the
derivative portal, we choose
\begin{equation}
\begin{aligned}
m_D &= 10^{-5}\,\mathrm{eV},
\qquad
\mu = 10^{-10}\,\mathrm{eV},\\
\delta m &= 2\times10^{-10}\,\mathrm{eV},
\qquad
f_{\rm eff}=2.03\times10^{11}\,\mathrm{GeV}.
\end{aligned}
\label{eq:benchmark_framework}
\end{equation}

For the periodic potential in
Eq.~\eqref{eq:majoron_potential},
the maximum potential energy density is
$V_{\max}=2m_\phi^2f_\phi^2
=2\times10^{-30}\,\mathrm{eV}^4$,
where $V_{\max}$ denotes the maximum
of the Majoron potential.
This is about $8\times10^{-20}$ of the
observed dark-energy density.
Since $\dot\phi(t_i)=0$, the initial
Majoron energy density satisfies
$\rho_\phi(t_i)\leq V_{\max}$.
The Majoron is therefore a subdominant
cosmological component, not the main
source of cosmic acceleration.

The collisionless calculation has three
distinct conditions.
First, the positive-energy two-state
description requires negligible mixing
with negative-energy states.
For the modes that dominate the response,
expansion-induced mixing is small when
$H\ll E_{\alpha,q}$.
Mixing induced by the derivative portal
must also remain small compared with
the corresponding energy gap.

Second, the linearized coherence equation
requires weak Majoron-induced transitions.
For nonzero $\Delta E_q$, a sufficient
small-mixing condition is
$|A_q\dot\phi/(f_{\rm eff}\Delta E_q)|\ll1$
for the relevant modes.
Within these approximations, the
finite-time retarded solution retains
the initial boundary transient.

Third, a local adiabatic expansion requires
additional conditions.
The Majoron source and $\Delta E_q$ must
vary slowly compared with $|\Delta E_q|$.
The initial boundary transient must be
treated separately.
In particular, $H\ll E_{\alpha,q}$
does not imply $H\ll|\Delta E_q|$.
Thus, the validity of the positive-energy
projection does not by itself justify
the local adiabatic approximation.

The exact Dirac limit also requires care.
At fixed observation time, $\delta m\to0$
must be taken in the finite-time retarded
solution. A static or long-time
approximation taken first can be
nonuniform as $\Delta E_q\to0$.
In the Dirac limit, the pseudo-Dirac
current is conserved.

\section{Spinor Projections and Current Response}
\label{sec:ward_current}

The effective action, exact operator Ward identity,
microscopic energy-transfer law, and initial state
were established in Sec.~\ref{sec:framework}.
We now calculate the vector-current and scalar-bilinear
matrix elements needed for the two-state density-matrix
description.

The positive-energy matrix elements are evaluated exactly at fixed comoving momentum and time.
However, restricting the dynamics to this subspace
is an approximation. In an expanding FRW universe,
the spinor basis changes with time and can mix
positive- and negative-energy states
\cite{Landete:2014}.
We first verify the algebraic Ward relation at fixed
time. We then examine the corrections caused by
the time-dependent basis and their effect on the
reduced description.

\subsection{Comoving modes and positive-energy spinors}
\label{subsec:ward_spinor_projection}

In the FRW background,
the rescaled Majorana field
$\psi_\alpha=a^{3/2}\Psi_\alpha$ removes the spin-connection
dilution term from the free mode equation. The rescaling does not remove the time dependence of $p=q/a$. For fixed helicity, the free two-component Hamiltonian is
\begin{equation}
\begin{aligned}
h_{\alpha,qh}(t)
&=\begin{pmatrix}
-hp&m_\alpha\\
m_\alpha&hp
\end{pmatrix}
=-hp\,\sigma_z+m_\alpha\sigma_x,\\
h_{\alpha,qh}^2&=E_{\alpha,q}^2 I_2,
\end{aligned}
\label{eq:III_hi}
\end{equation}
where $h_{\alpha,qh}$ acts on the reduced spinor space,
$\sigma_x$ and $\sigma_z$ are Pauli matrices, and $I_2$
is the $2\times2$ identity matrix.

For $m_\alpha>0$, a real positive-energy eigenvector is
\begin{equation}
\begin{aligned}
u_{\alpha,qh}
&=\begin{pmatrix}
\sqrt{\dfrac{E_{\alpha,q}-hp}{2E_{\alpha,q}}}\\[4pt]
\sqrt{\dfrac{E_{\alpha,q}+hp}{2E_{\alpha,q}}}
\end{pmatrix},\\
h_{\alpha,qh}u_{\alpha,qh}&=E_{\alpha,q}u_{\alpha,qh},
\qquad u_{\alpha,qh}^\dagger u_{\alpha,qh}=1,
\end{aligned}
\label{eq:III_ui}
\end{equation}
where $u_{\alpha,qh}$ is the normalized positive-energy
spinor of species $\alpha$.

In the combined Majorana-species and reduced-spinor space,
the free Hamiltonian is
\begin{equation}
\mathsf H_{0,qh}=
\begin{pmatrix}
h_{1,qh}&0\\
0&h_{2,qh}
\end{pmatrix}.
\label{eq:III_H4}
\end{equation}
The current matrix $\mathsf Q$ introduced in
Eq.~\eqref{eq:A_projection_framework} and the
scalar-bilinear matrix $\mathsf S$ are
\begin{equation}
\mathsf Q=\sigma_y\otimes I_2,
\qquad
\mathsf S=\sigma_x\otimes\sigma_x,
\label{eq:III_QS}
\end{equation}
where the first factor acts on Majorana-species space
and the second on reduced-spinor space; $\otimes$
denotes the tensor product. These matrices act in a
four-dimensional one-particle space containing both
energy signs.

Since $h_{1,qh}-h_{2,qh}=\delta m\,\sigma_x$,
direct matrix multiplication gives
\begin{equation}
i[\mathsf H_{0,qh},\mathsf Q]
=\delta m(\sigma_x\otimes\sigma_x)
=\delta m\,\mathsf S.
\label{eq:III_ward_matrix_exact}
\end{equation}
The derivative interaction adds
$-\dot\phi\,\mathsf Q/f_{\rm eff}$ to the one-particle
Hamiltonian. This term commutes with $\mathsf Q$ and
does not change Eq.~\eqref{eq:III_ward_matrix_exact}.
The equality holds instantaneously even though $p$
changes with time. It checks the sign of the operator
Ward identity in Eq.~\eqref{eq:III_ward_operator}; it
does not, by itself, justify dropping negative-energy
states during FRW evolution.

The normalized positive-energy states are
$|1,qh,+\rangle=(u_{1,qh},0)^T$ and
$|2,qh,+\rangle=(0,u_{2,qh})^T$.
They form the columns of $U_{qh}$ introduced in
Sec.~\ref{subsec:ensemble_scope}.
Its current projection is already given in
Eq.~\eqref{eq:A_projection_framework}.
The scalar-bilinear projection is
\begin{equation}
U_{qh}^\dagger\mathsf S U_{qh}=B_q\sigma_x.
\label{eq:III_AB_projection}
\end{equation}
The current overlap obeys
\begin{equation}
u_{1,qh}^\dagger u_{2,qh}=A_q,
\label{eq:III_A}
\end{equation}
whose explicit form is given once in
Eq.~\eqref{eq:A_exact_framework}.
The scalar overlap is
\begin{equation}
B_q\equiv u_{1,qh}^\dagger\sigma_x u_{2,qh}
=\left[
\frac{E_{1,q}E_{2,q}-p^2+m_1m_2}
{2E_{1,q}E_{2,q}}
\right]^{1/2},
\label{eq:III_B}
\end{equation}
where $B_q$ is the positive-energy matrix element of
the scalar bilinear. Both overlaps are independent of
helicity. They follow from the same normalized spinors;
$B_q$ is not an independent coupling.

The eigenvalue equations in
Eq.~\eqref{eq:III_ui} imply that the left side
of the following identity also equals
$\Delta E_q A_q$:
\begin{equation}
u_{1,qh}^\dagger(h_{1,qh}-h_{2,qh})u_{2,qh}
=\delta m\,u_{1,qh}^\dagger\sigma_xu_{2,qh}.
\label{eq:III_overlap_proof}
\end{equation}
Hence the exact fixed-time projected Ward relation is
\begin{equation}
\Delta E_q A_q=\delta m B_q.
\label{eq:III_AB_Ward}
\end{equation}
The equivalent two-state commutator check is
\begin{equation}
i[\operatorname{diag}(E_{1,q},E_{2,q}),A_q\sigma_y]
=\delta m B_q\sigma_x.
\label{eq:III_projected_commutator}
\end{equation}
Here $\operatorname{diag}$ denotes a diagonal matrix.
These identities fix the relative sign and normalization
at each time. They do not by themselves establish the
full time-dependent projected Ward equation.

For $\delta m\ll m_D$, the leading expansions are
\begin{align}
A_q&=1-\frac{(\delta m)^2p^2}
{8(p^2+m_D^2)^2}
+O\!\left(\frac{(\delta m)^4}{m_D^4}\right),
\label{eq:III_A_expansion}\\
B_q&=\frac{m_D}{\sqrt{p^2+m_D^2}}
+O\!\left(\frac{(\delta m)^2}{m_D^2}\right),
\label{eq:III_B_expansion}\\
\Delta E_q&=\frac{m_D\delta m}{\sqrt{p^2+m_D^2}}
+O\!\left(\frac{(\delta m)^3}{m_D^2}\right).
\label{eq:III_deltaE_expansion}
\end{align}
The largest leading deviation from $A_q=1$ occurs
at $p=m_D$:
\begin{equation}
\max_p(1-A_q)\simeq
\frac{(\delta m)^2}{32m_D^2}
=1.25\times10^{-11}.
\label{eq:III_A_benchmark}
\end{equation}
For the benchmark in Eq.~\eqref{eq:benchmark_framework},
the current overlap is therefore very close to unity.
The scalar overlap $B_q$, however, can be much smaller
than unity for relativistic modes and must be retained.

\subsection{Expectation values in the two-state density matrix}
\label{subsec:ward_rho_and_current}

We now express the projected current and scalar bilinear
in terms of the two-state density matrix. At each comoving
momentum and helicity, we use
\begin{equation}
\begin{aligned}
\rho_{kl,qh}(t)
&\equiv\langle a_{l,qh}^{\dagger}(t)a_{k,qh}(t)\rangle,\\
\rho_{qh}(t)
&=\begin{pmatrix}
f_{1,qh}(t)&\rho_{12,qh}(t)\\
\rho_{12,qh}^{*}(t)&f_{2,qh}(t)
\end{pmatrix}.
\end{aligned}
\label{eq:III_rho_convention}
\end{equation}
Here $a_{k,qh}$ annihilates a positive-energy Majorana
mode of species $k=1,2$, and $l=1,2$ is another species
index. The dagger denotes Hermitian conjugation.
These operators must not be confused with the scale
factor $a(t)$. The initial value of $\rho_{qh}$ is given
in Eq.~\eqref{eq:initial_rho_framework}.

With this index convention, the Pauli-matrix traces are
\begin{equation}
\begin{aligned}
\operatorname{tr}(\sigma_y\rho_{qh})
&=-2\operatorname{Im}\rho_{12,qh},\\
\operatorname{tr}(\sigma_x\rho_{qh})
&= 2\operatorname{Re}\rho_{12,qh}.
\end{aligned}
\label{eq:III_traces}
\end{equation}
Here $\operatorname{tr}$ denotes the trace over the two
Majorana species, while $\operatorname{Re}$ and
$\operatorname{Im}$ denote real and imaginary parts.
The signs follow from the density-matrix definition
in Eq.~\eqref{eq:III_rho_convention}.

Using the current projection in
Eq.~\eqref{eq:A_projection_framework}, we obtain
\begin{equation}
\langle J_{\rm PD}^0(t)\rangle_+
=-\frac{2}{a^3(t)}\sum_{h=\pm1}
\int\!\frac{d^3q}{(2\pi)^3}\,
A_q(t)\operatorname{Im}\rho_{12,qh}(t).
\label{eq:III_J_density_matrix}
\end{equation}
The corresponding scalar-bilinear expectation value is
\begin{equation}
\langle:\bar\Psi_1\Psi_2:\rangle_+
=\frac{2}{a^3(t)}\sum_{h=\pm1}
\int\!\frac{d^3q}{(2\pi)^3}\,
B_q(t)\operatorname{Re}\rho_{12,qh}(t).
\label{eq:III_scalar_density_matrix}
\end{equation}
Here the subscript $+$ denotes the positive-energy
one-particle projection. The factor $a^{-3}$ converts
the comoving phase-space measure into a physical density.
The factor of two in each expression comes from the
Pauli-matrix trace in Eq.~\eqref{eq:III_traces}.
The helicity sum is already explicit and must not be
counted again. Normal ordering subtracts the chosen
source-free reference-vacuum contribution. Pair coherence and negative-energy contributions
are not included in these projected expectation values.
They belong to the full field-theory description
\cite{Blaschke:2016xxt}.

The Hamiltonian governing these positive-energy states
in the homogeneous Majoron background is
\begin{equation}
H_{qh}^{(+)}(t)=
\begin{pmatrix}E_{1,q}(t)&0\\0&E_{2,q}(t)\end{pmatrix}
-\frac{A_q(t)\dot\phi(t)}{f_{\rm eff}}\sigma_y.
\label{eq:III_Hpositive}
\end{equation}
This is the same two-state Hamiltonian used in the
initial-condition discussion in
Sec.~\ref{subsec:ensemble_scope}. We neglect
nonadiabatic mixing with negative-energy states here.
The collisionless Liouville equation
$\dot\rho_{qh}=-i[H_{qh}^{(+)},\rho_{qh}]$ then gives,
to first order in the derivative drive,
\begin{equation}
\dot\rho_{12,qh}^{(1)}
+i\Delta E_q(t)\rho_{12,qh}^{(1)}
=\frac{A_q(t)D_q^{(0)}}{f_{\rm eff}}\dot\phi(t).
\label{eq:III_rho12_bridge}
\end{equation}
Here the superscript $(1)$ denotes the first-order
coherence. The zeroth-order occupation difference
$D_q^{(0)}$ and the initial condition are specified in
Eqs.~\eqref{eq:free_streaming_framework} and
\eqref{eq:initial_rho_framework}.

The source on the right-hand side is real, while
$\Delta E_q$ sets the phase evolution. Corrections
to the diagonal occupations begin at second order
in the drive. Their feedback on the coherence
therefore enters at third order within the closed
two-state dynamics. Below, we omit the superscript
$(1)$ when discussing the linear response.

\subsection{Time-dependent projections versus the exact Ward identity}
\label{subsec:ward_time_basis}

The fixed-time relation in Eq.~\eqref{eq:III_AB_Ward}
does not ensure that the truncated positive-energy
dynamics satisfies the full FRW Ward identity.
The positive-energy projector
$P_{qh}(t)=U_{qh}(t)U_{qh}^{\dagger}(t)$
changes as the physical momentum $p=q/a(t)$ redshifts.
The projected current overlap $A_q(t)$ therefore
also changes with time.

Differentiating Eq.~\eqref{eq:III_J_density_matrix}
and using the linearized equation
\eqref{eq:III_rho12_bridge} gives, within the
instantaneous positive-energy approximation,
\begin{align}
&\left(\partial_t+3H\right)
\langle J_{\rm PD}^0\rangle_+\nonumber\\
&\quad=\frac{2}{a^3}\sum_h
\int\!\frac{d^3q}{(2\pi)^3}
\left[
A_q\Delta E_q\operatorname{Re}\rho_{12,qh}
-\dot A_q\operatorname{Im}\rho_{12,qh}
\right].
\label{eq:III_projected_divergence}
\end{align}
The first term is $\delta m$ times the projected
scalar-bilinear expectation value in
Eq.~\eqref{eq:III_scalar_density_matrix}, by
Eq.~\eqref{eq:III_AB_Ward}. The second term
arises from the time dependence of the projected
current operator. It shows why the truncated
relation cannot be identified with the exact
operator Ward identity.

To track the omitted basis contributions, we use
the complete four-dimensional one-particle space.
The $4\times2$ matrix $U_{qh}$ contains only the
positive-energy states. We distinguish it from the
$4\times4$ unitary matrix
$\mathcal U_{qh}=(U_{qh},U_{qh}^{(-)})$, where
$U_{qh}^{(-)}$ contains the two normalized
negative-energy eigenstates of the free Hamiltonian
$\mathsf H_{0,qh}$. The full one-particle Hamiltonian is
$\mathsf H_{qh}=\mathsf H_{0,qh}
-(\dot\phi/f_{\rm eff})\mathsf Q$.
In the complete instantaneous free-energy basis, define
\begin{equation}
\begin{aligned}
\Omega_{qh}
&=\mathcal U_{qh}^{\dagger}\dot{\mathcal U}_{qh},\\
\mathsf Q_e
&=\mathcal U_{qh}^{\dagger}\mathsf Q\mathcal U_{qh},\\
\mathsf H_e
&=\mathcal U_{qh}^{\dagger}\mathsf H_{qh}\mathcal U_{qh}
-i\Omega_{qh}.
\end{aligned}
\label{eq:III_eigenbasis_def}
\end{equation}
Here $\Omega_{qh}$ is the geometric basis connection,
and the subscript $e$ denotes matrices expressed in
this complete instantaneous basis.

The canonical-basis matrix $\mathsf Q$ has no explicit
time dependence. Unitarity of $\mathcal U_{qh}$ and
the full matrix Ward relation
\eqref{eq:III_ward_matrix_exact} therefore give
\begin{equation}
\begin{aligned}
\dot{\mathsf Q}_e
&=[\mathsf Q_e,\Omega_{qh}],\\
\dot{\mathsf Q}_e+i[\mathsf H_e,\mathsf Q_e]
&=i[\mathcal U_{qh}^{\dagger}\mathsf H_{qh}
\mathcal U_{qh},\mathsf Q_e]\\
&=\delta m\,\mathcal U_{qh}^{\dagger}
\mathsf S\mathcal U_{qh}.
\end{aligned}
\label{eq:III_covariant_eigenbasis_Ward}
\end{equation}
The geometric connection in $\mathsf H_e$ cancels
the explicit basis derivative of $\mathsf Q_e$.
Taking the full expectation value and restoring
the common $a^{-3}$ dilution is consistent with
the FRW Ward identity in Eq.~\eqref{eq:III_ward_FRW}.
The cancellation does not imply charge conservation:
for nonzero $\delta m$, the scalar-bilinear source
remains. This is a complete-basis statement, not an
exact identity for the truncated positive-energy current.
In particular, negative-energy and pair-coherence
contributions omitted from
Eq.~\eqref{eq:III_projected_divergence}
can be needed for the full cancellation.
Such coherence sectors arise in more general
fermionic quantum-kinetic descriptions
\cite{Herranen:2009coherence}. The residual
$\dot A_q$ term must be assessed as a projection
error, not interpreted as irreversible dissipation.

The change of the free spinor basis produces a
positive/negative-energy connection. For a normalized
negative-energy eigenvector $v_{\alpha,qh}$ orthogonal
to $u_{\alpha,qh}$, its magnitude is
\begin{equation}
\left|\langle v_{\alpha,qh}|\dot u_{\alpha,qh}\rangle\right|
=\frac{Hm_\alpha p}{2E_{\alpha,q}^2}.
\label{eq:III_nonadiabatic_connection}
\end{equation}
The corresponding dimensionless parameter is
$\epsilon_{\alpha,q}^{\rm geom}
\equiv |\langle v_{\alpha,qh}|\dot u_{\alpha,qh}\rangle|
/(2E_{\alpha,q})
=Hm_\alpha p/(4E_{\alpha,q}^3)
\le H/(8E_{\alpha,q})$.
Thus $H\ll E_{\alpha,q}$ suppresses the geometric
mixing for the modes contributing to the response.

The size of the overlap variation follows from
Eq.~\eqref{eq:III_A_expansion} and $\dot p=-Hp$:
\begin{align*}
\frac{\dot A_q}{A_q}
&=\frac{H(\delta m)^2p^2(m_D^2-p^2)}
{4(p^2+m_D^2)^3}\\
&\quad+O\!\left[H\left(\frac{\delta m}{m_D}\right)^4\right].
\end{align*}
At leading order, maximizing the absolute value
over $p\geq0$ gives
$|\dot A_q/A_q|\le
(\sqrt{3}/72)H(\delta m/m_D)^2$.
For Eq.~\eqref{eq:benchmark_framework}, this is
$9.62\times10^{-12}H$.
This bounds the relative variation of $A_q$;
it does not give a uniform relative bound on the
Ward source, which can pass through zero.

The derivative interaction can also couple positive-
and negative-energy states. Neglecting this mixing
requires, for the relevant modes,
$(|\dot\phi|/f_{\rm eff})
|u_{1,qh}^{\dagger}v_{2,qh}|
\ll E_{1,q}+E_{2,q}$.
This condition is distinct from the geometric
suppression estimated above. Both control the
positive-energy approximation. Neither alters
the exact field-theoretic Ward identity.

\subsection{Microscopic Majoron energy transfer}
\label{subsec:ward_energy_exchange}

The Majoron continuity equation and the exact microscopic
energy-transfer relation were derived in
Eqs.~\eqref{eq:phi_continuity_framework} and
\eqref{eq:Q_ward_framework}. Their derivation uses the
full operator Ward identity, not a positive-energy
projection. We now examine how the source is evaluated
within the reduced two-state description.

Substituting the projected scalar-bilinear expectation
value from Eq.~\eqref{eq:III_scalar_density_matrix}
into the Ward source gives
\begin{equation}
Q_{\rm micro}^{(+)}(t)
=\frac{2\delta m\,\dot\phi(t)}{a^3(t)f_{\rm eff}}
\sum_h\int\!\frac{d^3q}{(2\pi)^3}\,
B_q(t)\operatorname{Re}\rho_{12,qh}(t).
\label{eq:III_Q_projected}
\end{equation}
This is the positive-energy estimate of the microscopic
source. Its validity is restricted by the projection
conditions discussed in
Sec.~\ref{subsec:ward_time_basis}.
It does not require a local adiabatic expansion
in $\Delta E_q$.
However, the projected scalar source and the
divergence of the projected current need not agree
exactly in a time-dependent spinor basis.
Their difference is evaluated below.

The exact source can be separated into the projected
contribution and the omitted matrix elements:
\begin{equation}
\begin{aligned}
Q_{\rm micro}(t)
&=Q_{\rm micro}^{(+)}(t)\\
&\quad+\frac{\delta m\,\dot\phi(t)}{f_{\rm eff}}
\left[
\left\langle:\bar\Psi_1\Psi_2:\right\rangle
-\left\langle:\bar\Psi_1\Psi_2:\right\rangle_+
\right].
\end{aligned}
\label{eq:III_Q_exact}
\end{equation}
This identity follows from
Eq.~\eqref{eq:Q_ward_framework}. It makes the
projection limitation explicit but does not, by
itself, bound the omitted contribution.

A related distinction arises if the transfer rate is
instead inferred by differentiating the projected
current. Combining Eqs.~\eqref{eq:III_projected_divergence}
and \eqref{eq:III_Q_projected} gives
\begin{equation}
\begin{aligned}
&Q_{\rm micro}^{(+)}
-\frac{\dot\phi}{f_{\rm eff}}
(\partial_t+3H)\langle J_{\rm PD}^0\rangle_+\\
&\quad=\frac{2\dot\phi}{a^3 f_{\rm eff}}
\sum_h\int\!\frac{d^3q}{(2\pi)^3}\,
\dot A_q\operatorname{Im}\rho_{12,qh}.
\end{aligned}
\label{eq:III_energy_continuity}
\end{equation}
The right-hand side is caused by the time dependence
of the projected current matrix element. It need not
vanish within the truncated dynamics, even though the
full operator Ward identity remains exact. Neither this
projection residual nor the omitted contribution in
Eq.~\eqref{eq:III_Q_exact} should be interpreted as
irreversible dissipation.

The Dirac limit requires a finite-time treatment.
For the prepared ensemble,
Eq.~\eqref{eq:occupation} gives
$D_q(t_i)=O[\Delta E_q(t_i)/T_{\rm PD,i}]$.
At fixed comoving momentum, this difference vanishes
with $\delta m$. For a bounded Majoron drive and any
fixed finite time interval, the solution of the
linearized initial-value equation
\eqref{eq:III_rho12_bridge}, with
$\rho_{12,qh}(t_i)=0$, remains regular as
$\Delta E_q\to0$. In particular, the induced coherence
vanishes with $D_q(t_i)$, and the projected source in
Eq.~\eqref{eq:III_Q_projected} is of order
$(\delta m)^2$ at fixed finite time.

A formal zero-frequency particular solution can contain
inverse powers of $\Delta E_q$. Its homogeneous
counterpart is not damped in the collisionless theory
and is required to enforce the prepared initial condition.
Consequently, taking the long-time or static limit before
the Dirac limit can give a nonuniform result. A local
static approximation must not be extrapolated into
$|\Delta E_q|(t-t_i)\lesssim1$.
In the exact Dirac limit the full current is conserved,
as required by Eq.~\eqref{eq:III_ward_FRW}.

The fixed-time spinor overlaps, their relative signs,
and the modewise Ward relation
\eqref{eq:III_AB_Ward} thus provide the microscopic
input for the retarded Majoron response, while
Eqs.~\eqref{eq:III_Q_exact} and
\eqref{eq:III_energy_continuity} distinguish the
exact source from its reduced representation.

\section{Coherent Quantum Kinetics}
\label{sec:coherent_kinetics}

We now solve the collisionless initial-value problem
specified in Sec.~\ref{subsec:ensemble_scope},
using the two-state coherence equation and current
matrix elements derived in Sec.~\ref{sec:ward_current}.

The solution retains the relative phase accumulated
between the two Majorana mass eigenstates at fixed
comoving momentum. It also preserves the finite-time
initial condition and the resulting boundary transient.
The induced current is therefore described by a
retarded memory integral rather than an assumed
local friction law.

This integral is the exact solution of the linearized
instantaneous positive-energy two-state equation.
Its validity remains subject to the collisionless
evolution and positive-energy projection
approximations established above.
It is not an exact solution of the complete,
unprojected field-theoretic dynamics.

\subsection{Collisionless initial-value problem}
\label{subsec:IV_initial_value}

The mode energies and their splitting were defined in
Eq.~\eqref{eq:energies_framework}. At fixed comoving momentum, their time dependence follows from $\dot p=-Hp$:
\begin{equation}
\begin{aligned}
\dot E_{\alpha,q}&=-\frac{Hp^2}{E_{\alpha,q}},\\
\dot{\Delta E}_q&=\frac{Hp^2\Delta E_q}{E_{1,q}E_{2,q}}.
\end{aligned}
\label{eq:IV_energy}
\end{equation}
The level splitting is therefore time dependent even though
the masses are constant.

We use the density-matrix convention of
Eq.~\eqref{eq:III_rho_convention} and introduce the shorthand
\begin{equation}
\begin{aligned}
r_{qh}(t)&\equiv\rho_{12,qh}(t),\\
D_{qh}(t)&\equiv f_{2,qh}(t)-f_{1,qh}(t).
\end{aligned}
\label{eq:IV_rho}
\end{equation}
For the prepared state in
Eqs.~\eqref{eq:initial_rho_framework}--\eqref{eq:occupation},
\begin{equation}
r_{qh}(t_i)=0,\qquad
D_{qh,i}\equiv D_{qh}(t_i)=D_q(t_i)>0.
\label{eq:IV_initial}
\end{equation}
The positivity of the prepared difference also follows
without expanding in the splitting:
\begin{equation}
D_{qh,i}=\int_{E_{2,q}(t_i)/T_{\rm PD,i}}^{E_{1,q}(t_i)/T_{\rm PD,i}}
 dx\,n_F(x)[1-n_F(x)]>0.
\label{eq:IV_initial_D}
\end{equation}
Its small-splitting expansion is already given in
Eq.~\eqref{eq:occupation}. In particular, this difference
is fixed by the prepared ensemble, not by a chosen
hidden charge asymmetry.

For the subsequent evolution, separate the prepared
occupations from their induced changes:
\begin{equation}
\begin{aligned}
f_{\alpha,qh}(t)&=f_{\alpha,q}(t_i)
 +\delta f_{\alpha,qh}(t),\\
D_{qh}(t)&=D_{qh,i}+\delta D_{qh}(t).
\end{aligned}
\label{eq:IV_free}
\end{equation}
Here $\delta f_{\alpha,qh}$ and $\delta D_{qh}$ are the
changes from the prepared values. At zeroth order they
vanish, as stated in q.~\eqref{eq:free_streaming_framework}.
No instantaneous thermal distribution is imposed after
$t_i$ \cite{BoyanovskyHo2004}. Possible pre-preparation interactions do not enter the collisionless evolution. General quantum-kinetic formulations are discussed in
Refs.~\cite{Sigl:1993ct,Vlasenko:2013fja}.

The two-state Hamiltonian is the same $H_{qh}^{(+)}$
as in Eq.~\eqref{eq:III_Hpositive}. Its identity-matrix
part does not affect the density-matrix evolution. Thus
we use the traceless Hamiltonian
\begin{equation}
\begin{aligned}
\widetilde H_{qh}^{(+)}(t)
&\equiv H_{qh}^{(+)}(t)
 -\frac{E_{1,q}(t)+E_{2,q}(t)}{2}I_2\\
&=\frac{\Delta E_q(t)}{2}\sigma_z-g_q(t)\sigma_y,\\
g_q(t)&\equiv\frac{A_q(t)\dot\phi(t)}{f_{\rm eff}}.
\end{aligned}
\label{eq:IV_H}
\end{equation}
Here $g_q$ is the Majoron-induced off-diagonal drive.
Within the instantaneous positive-energy two-state
approximation, the collisionless equation is
\begin{equation}
\dot\rho_{qh}=-i[\widetilde H_{qh}^{(+)},\rho_{qh}].
\label{eq:IV_liouville}
\end{equation}
Its matrix components give
\begin{align}
\dot r_{qh}+i\Delta E_q r_{qh}
 &=g_q D_{qh}(t),
\label{eq:IV_full_coherence}\\
\dot f_{1,qh}&=2g_q\operatorname{Re}r_{qh},\qquad
\dot f_{2,qh}=-2g_q\operatorname{Re}r_{qh},
\label{eq:IV_diagonal_evolution}\\
\dot D_{qh}&=-4g_q\operatorname{Re}r_{qh}.
\label{eq:IV_D_evolution}
\end{align}
They conserve the trace $f_{1,qh}+f_{2,qh}$ and
preserve the eigenvalues of the reduced density matrix.
With the prepared initial condition this gives
\begin{equation}
D_{qh}^2(t)+4|r_{qh}(t)|^2=D_{qh,i}^2.
\label{eq:IV_bloch_invariant}
\end{equation}
This invariant is exact within the closed, unitary
two-state model. The conserved trace is not the
pseudo-Dirac charge, which is associated with the
off-diagonal current and need not be conserved for
nonzero mass splitting.

To interpret the invariant, let $\mathcal W_{qh}(t,t_i)$
be the two-state evolution operator after removal of an
irrelevant overall phase. It must not be confused with
the positive-energy spinor matrix $U_{qh}$ or the complete
spinor-basis matrix $\mathcal U_{qh}$.
The unitary evolution of the prepared density matrix is
\[
\rho_{qh}(t)=\mathcal W_{qh}(t,t_i)\rho_{qh}(t_i)
\mathcal W_{qh}^{\dagger}(t,t_i).
\]
Writing
\[
\mathcal W_{qh}=\begin{pmatrix}
c_{qh}&s_{qh}\\-s_{qh}^{*}&c_{qh}^{*}
\end{pmatrix},\qquad
|c_{qh}|^2+|s_{qh}|^2=1,
\]
and defining the transition probability
$\mathcal P_{qh}(t)\equiv|s_{qh}(t)|^2$, we obtain
\[
\begin{aligned}
D_{qh}(t)&=D_{qh,i}[1-2\mathcal P_{qh}(t)],\\
|r_{qh}(t)|^2&=D_{qh,i}^2\mathcal P_{qh}(t)
 [1-\mathcal P_{qh}(t)].
\end{aligned}
\]
The notation $\mathcal P_{qh}$ distinguishes this
probability from the positive-energy projector $P_{qh}$
in Sec.~\ref{subsec:ward_time_basis}.

Equal diagonal occupations occur at
$\mathcal P_{qh}=1/2$, when
$|r_{qh}|=D_{qh,i}/2$. This is the largest
coherence compatible with the unchanged eigenvalues
of the prepared density matrix. It is neither thermal
equilibration nor a maximally mixed state. Nor is it
generally stationary: Eq.~\eqref{eq:IV_D_evolution}
can remain nonzero when $D_{qh}=0$.
Reaching this condition requires an order-unity
transition probability and lies beyond the linear
response used below, even if the instantaneous drive
is small.

Starting from $r_{qh}(t_i)=0$, the induced coherence
is first order in the derivative interaction, while
$\delta f_{\alpha,qh}$ and $\delta D_{qh}$ begin at
second order. Their feedback into
Eq.~\eqref{eq:IV_full_coherence} enters at third order.
For a prescribed Majoron background, $r_{qh}^{(1)}$
is of order $f_{\rm eff}^{-1}$, so the leading
Majoron energy transfer is of order $f_{\rm eff}^{-2}$.
It is therefore consistent at this order to keep
$D_{qh,i}$ fixed in the source. The first-order
coherence obeys Eq.~\eqref{eq:III_rho12_bridge}.
For its finite-time integration, define the integrating
factor $\mathcal F_q(t)\equiv
\exp[i\int_{t_i}^{t}ds\,\Delta E_q(s)]$.
The linear initial-value equation can then be written
without repeating Eq.~\eqref{eq:III_rho12_bridge}:
\begin{equation}
\begin{aligned}
\frac{d}{dt}\bigl[\mathcal F_q(t)r_{qh}^{(1)}(t)\bigr]
 &=\mathcal F_q(t)g_q(t)D_{qh,i},\\
r_{qh}^{(1)}(t_i)&=0.
\end{aligned}
\label{eq:IV_linear_eq}
\end{equation}
The phase and source signs are fixed by the same
Hamiltonian and current conventions used in Sec.~\ref{sec:ward_current}.
No collision or dephasing term is added. Corrections
from the time-dependent positive/negative-energy basis
remain separate from this closed two-state evolution.

\subsection{Exact finite-time retarded solution}
\label{subsec:IV_exact_solution}

The linearized initial-value equation has already been
written in integrating-factor form in
Eq.~\eqref{eq:IV_linear_eq}. To express its solution,
we define the phase accumulated between $t'$ and $t$:
\begin{equation}
\Theta_q(t,t')\equiv\int_{t'}^t ds\,\Delta E_q(s).
\label{eq:IV_phase}
\end{equation}
Integrating Eq.~\eqref{eq:IV_linear_eq} from the
preparation time $t_i$ gives
\begin{equation}
r_{qh}(t)=\frac{D_{qh,i}}{f_{\rm eff}}
\int_{t_i}^{t}dt'\,A_q(t')\dot\phi(t')
e^{-i\Theta_q(t,t')}.
\label{eq:IV_retarded_solution}
\end{equation}
This is the exact retarded solution of the linearized,
collisionless positive-energy two-state equation, not
of the unprojected field theory. Its finite lower limit
preserves the prepared initial condition and the
associated boundary transient. No damping or
phenomenological friction term has been introduced.

To display its phase and sign conventions, write the
real and imaginary parts as
\begin{equation}
\begin{aligned}
\operatorname{Re}r_{qh}(t)
&=\frac{D_{qh,i}}{f_{\rm eff}}I_{c,qh}(t),\\
I_{c,qh}(t)
&\equiv\int_{t_i}^{t}dt'\,A_q(t')\dot\phi(t')
\cos\Theta_q(t,t').
\end{aligned}
\label{eq:IV_real_r}
\end{equation}
\begin{equation}
\begin{aligned}
\operatorname{Im}r_{qh}(t)
&=-\frac{D_{qh,i}}{f_{\rm eff}}I_{s,qh}(t),\\
I_{s,qh}(t)
&\equiv\int_{t_i}^{t}dt'\,A_q(t')\dot\phi(t')
\sin\Theta_q(t,t').
\end{aligned}
\label{eq:IV_imag_r}
\end{equation}
The minus sign in the imaginary component is fixed
by the phase in Eq.~\eqref{eq:IV_retarded_solution}.

Using Eq.~\eqref{eq:III_J_density_matrix}, the induced
current is
\begin{equation}
\begin{aligned}
\langle J_{\rm PD}^{0}(t)\rangle_+
=\frac{2}{a^3(t)f_{\rm eff}}
\sum_h\int\!\frac{d^3q}{(2\pi)^3}\,
A_q(t)D_{qh,i}I_{s,qh}(t).
\end{aligned}
\label{eq:IV_J_nonlocal}
\end{equation}
Equations~\eqref{eq:IV_retarded_solution} and
\eqref{eq:IV_J_nonlocal} give the finite-time
coherence and current response used below.
The scalar-bilinear projection in
Eq.~\eqref{eq:III_scalar_density_matrix} similarly gives
\begin{equation}
\langle:\bar\Psi_1\Psi_2:\rangle_+
=\frac{2}{a^3(t)f_{\rm eff}}
\sum_h\int\!\frac{d^3q}{(2\pi)^3}\,
B_q(t)D_{qh,i}I_{c,qh}(t).
\label{eq:IV_scalar_nonlocal}
\end{equation}
The physical-density factor and explicit helicity sum
have the normalization established in
Sec.~\ref{subsec:ward_rho_and_current}; no additional
spin-degeneracy factor is introduced.

Inserting Eq.~\eqref{eq:IV_scalar_nonlocal} into the
exact Ward-based source
\eqref{eq:Q_ward_framework} gives its leading
positive-energy estimate:
\begin{equation}
\begin{aligned}
Q_{\rm micro}^{(+),W}(t)
=\frac{2\delta m\,\dot\phi(t)}{a^3(t)f_{\rm eff}^2}
\sum_h\int\!\frac{d^3q}{(2\pi)^3}\,
B_q(t)D_{qh,i}I_{c,qh}(t).
\end{aligned}
\label{eq:IV_Q_W_memory}
\end{equation}
The superscript $W$ identifies evaluation through
the Ward scalar-bilinear source. Thus
$Q_{\rm micro}^{(+),W}=Q_{\rm micro}^{(+)}$ as
introduced in Eq.~\eqref{eq:III_Q_projected}; it
is not a separate transfer mechanism. The exact
field-theoretic source remains
Eq.~\eqref{eq:Q_ward_framework}, with the
projection limitation made explicit in
Eq.~\eqref{eq:III_Q_exact}. In particular, the
sign of the memory integral is not restricted to
that of a positive-definite friction term.

An independent consistency check follows by
differentiating the projected current. The memory
integrals obey
$\dot I_{s,qh}=\Delta E_q I_{c,qh}$:
the upper-limit term vanishes because
$\sin\Theta_q(t,t)=0$. Consequently,
\begin{equation}
\begin{aligned}
&(\partial_t+3H)\langle J_{\rm PD}^{0}\rangle_+\\
&\quad=\frac{2}{a^3 f_{\rm eff}}
\sum_h\int\!\frac{d^3q}{(2\pi)^3}\,
D_{qh,i}\bigl[
A_q\Delta E_q I_{c,qh}
+\dot A_q I_{s,qh}\bigr].
\end{aligned}
\label{eq:IV_current_divergence}
\end{equation}
By the fixed-time relation
\eqref{eq:III_AB_Ward}, the first term is
$\delta m\langle:\bar\Psi_1\Psi_2:\rangle_+$.
The second term reproduces the time-dependent
projection residual of
Eq.~\eqref{eq:III_energy_continuity} upon using
Eq.~\eqref{eq:IV_imag_r}. It does not signal a
violation of the full operator Ward identity,
which remains valid with the complete spinor basis
and its geometric connection, as established in
Sec.~\ref{subsec:ward_time_basis}. We do not
discard this residual or interpret it as
irreversible dissipation.

For numerical evolution, introduce the e-fold
variable $\mathcal N\equiv\ln a$, with
$\mathcal N_i\equiv\ln a(t_i)$. The phase and
retarded solution then become
\begin{equation}
\Theta_q(\mathcal N,\mathcal N')
=\int_{\mathcal N'}^{\mathcal N}d\bar{\mathcal N}\,
\frac{\Delta E_q(\bar{\mathcal N})}
{H(\bar{\mathcal N})},
\label{eq:IV_phase_efold}
\end{equation}
\begin{equation}
\begin{aligned}
r_{qh}(\mathcal N)
&=\frac{D_{qh,i}}{f_{\rm eff}}
\int_{\mathcal N_i}^{\mathcal N}d\mathcal N'\,
A_q(\mathcal N')\\
&\qquad\times\frac{d\phi}{d\mathcal N'}
e^{-i\Theta_q(\mathcal N,\mathcal N')}.
\end{aligned}
\label{eq:IV_solution_efold}
\end{equation}
The barred variable $\bar{\mathcal N}$ is the
dummy integration variable in the phase.
Since $dt=d\mathcal N/H$ and
$\dot\phi=H\,d\phi/d\mathcal N$, their factors of
$H$ cancel in the source integral. The momentum
integration continues to use conserved comoving $q$.

\subsection{Transient coherence and the absence of assumed friction}
\label{subsec:IV_transient}

To illustrate the finite-time response, consider an auxiliary sudden-switching test with constant coefficients. Neglect expansion
and hold $A_q$, $D_{qh,i}$, $\Delta E_q$, and $\dot\phi$ constant for $t>t_i$. The drive is switched on immediately after $t_i$;
this test does not impose the smooth cosmological condition
$\dot\phi(t_i)=0$. With $\tau\equiv t-t_i$ and the drive $g_q$ defined in Eq.~\eqref{eq:IV_H}, the retarded solution
\eqref{eq:IV_retarded_solution} becomes
\begin{equation}
r_{qh}(t)=\frac{g_qD_{qh,i}}{i\Delta E_q}
\left[1-e^{-i\Delta E_q\tau}\right].
\label{eq:IV_constant_example}
\end{equation}
Its real and imaginary parts are
\begin{equation}
\begin{aligned}
\operatorname{Re}r_{qh}(t)
&=\frac{g_qD_{qh,i}}{\Delta E_q}
\sin(\Delta E_q\tau),\\
\operatorname{Im}r_{qh}(t)
&=-\frac{g_qD_{qh,i}}{\Delta E_q}
\left[1-\cos(\Delta E_q\tau)\right].
\end{aligned}
\label{eq:IV_constant_components}
\end{equation}
The real part oscillates, while the imaginary part contains
a constant particular solution and an undamped oscillatory
term required by the initial condition. The projected current in Eq.~\eqref{eq:IV_J_nonlocal} consequently contains a constant offset and a persistent transient. The apparent inverse power
of $\Delta E_q$ in Eq.~\eqref{eq:IV_constant_example} is
regular at any fixed finite $\tau$ when the two terms are
retained together.

For constant $\dot\phi$, the Ward-based energy transfer
from Eq.~\eqref{eq:IV_Q_W_memory} is proportional, mode by
mode, to $\sin(\Delta E_q\tau)$. It changes sign during each cycle, and its average over a complete cycle vanishes.
There is no collision term to damp the coherent transient.
A time average or phase mixing across momenta can reduce
its visible oscillations, but neither is equivalent to
collision-induced entropy production. This example does
not justify a postulated positive-definite local friction law.

At fixed background coefficients, the retarded sine kernel
can also be represented in frequency space. Using the
Fourier convention $e^{-i\Omega t}$, we define the
frozen-background retarded susceptibility
$\chi_{\rm PD}^{R}(\Omega)$ through
$\langle J_{\rm PD}^{0}\rangle_+(\Omega)
=\chi_{\rm PD}^{R}(\Omega)\dot\phi(\Omega)/f_{\rm eff}$,
where $\Omega$ is the Fourier frequency. The causal kernel
in Eq.~\eqref{eq:IV_J_nonlocal} then gives
\begin{equation}
 \chi_{\rm PD}^{R}(\Omega)
 =\frac{2}{a^3}\sum_h\int\!\frac{d^3q}{(2\pi)^3}\,
 \frac{A_q^2D_{qh,i}\,\Delta E_q}
 {(\Delta E_q)^2-(\Omega+i0^+)^2}.
 \label{eq:IV_susceptibility_frequency}
\end{equation}
Here $0^+$ denotes a positive infinitesimal specifying
the retarded boundary condition, not a physical collision
or decoherence rate. The real part includes a dispersive
principal-value contribution. The imaginary part has
resonant spectral support at $|\Omega|=\Delta E_q$.
A sustained oscillatory drive may therefore transfer energy
through collisionless resonant response, while phase mixing
can suppress macroscopic coherence without collisions.
Neither effect alone establishes a local,
momentum-independent Markovian friction coefficient.

This frequency-domain expression assumes frozen background
coefficients. Its static limit cannot replace the finite-time, time-dependent solution without controlling the variation rates and the boundary transient. The corresponding local adiabatic conditions are examined in
Sec.~\ref{subsec:IV_adiabatic}.

\subsection{Adiabatic expansion and reactive susceptibility}
\label{subsec:IV_adiabatic}

We now obtain a local adiabatic expansion of the finite-time
retarded solution in Eq.~\eqref{eq:IV_retarded_solution},
following the general logic of adiabatic perturbation theory
\cite{RigolinOrtizPonce2008}.
For compactness, define
\begin{equation}
\begin{aligned}
\omega_q(t)&\equiv\Delta E_q(t),\\
F_{qh}(t)&\equiv g_q(t)D_{qh,i}\\
&=\frac{A_q(t)D_{qh,i}}{f_{\rm eff}}\dot\phi(t).
\end{aligned}
\label{eq:IV_omega_s}
\end{equation}
Here $F_{qh}$ is the real source of the linearized
coherence equation. It is distinct from the evolution-
operator amplitude $s_{qh}$ used in
Sec.~\ref{subsec:IV_initial_value}.

Using
$\partial_{t'}e^{-i\Theta_q(t,t')}
=i\omega_q(t')e^{-i\Theta_q(t,t')}$,
integration by parts gives the exact identity
\begin{align}
r_{qh}(t)={}&\frac{F_{qh}(t)}{i\omega_q(t)}
-\frac{F_{qh}(t_i)}{i\omega_q(t_i)}
 e^{-i\Theta_q(t,t_i)}\nonumber\\
&-\int_{t_i}^{t}dt'\,e^{-i\Theta_q(t,t')}
\frac{d}{dt'}\left[\frac{F_{qh}(t')}{i\omega_q(t')}\right].
\label{eq:IV_exact_parts}
\end{align}
The second term is the leading initial boundary
contribution. It vanishes for the prepared condition
$\dot\phi(t_i)=0$, but this does not remove the
higher-derivative boundary terms.

Repeated integration by parts gives the local
particular-solution series
\begin{align}
r_{qh}^{\rm ad}(t)={}&-i\frac{F_{qh}}{\omega_q}
+\frac{1}{\omega_q}\frac{d}{dt}
 \left(\frac{F_{qh}}{\omega_q}\right)\nonumber\\
&+\frac{i}{\omega_q}\frac{d}{dt}
 \left[\frac{1}{\omega_q}\frac{d}{dt}
 \left(\frac{F_{qh}}{\omega_q}\right)\right]
+\cdots.
\label{eq:IV_ad_expansion}
\end{align}
This is an asymptotic expansion of the local
particular solution, not an exact replacement for
the finite-time retarded solution. Its first term is
imaginary, its first derivative correction is real,
and the next term is imaginary. Thus the first local
derivative correction to the current begins at
second derivative order, rather than through a term
linear in $\ddot\phi$ in the current itself.

The boundary terms are essential near preparation.
Indeed, a second integration by parts produces, when
$F_{qh}(t_i)=0$, the leading surviving boundary term
\[
-e^{-i\Theta_q(t,t_i)}
\frac{\dot F_{qh}(t_i)}{\omega_q^2(t_i)}.
\]
Here $\dot F_{qh}(t_i)=
A_q(t_i)D_{qh,i}\ddot\phi(t_i)/f_{\rm eff}$.
This term cancels the first-derivative particular
solution at $t_i$, as required by $r_{qh}(t_i)=0$;
it also contributes to the initial time derivative
of the current. Dropping it before applying the
initial condition can therefore change the inferred
Majoron acceleration.

The adiabatic expansion requires, for modes that
contribute appreciably to the momentum integral,
\begin{equation}
\left|\frac{\dot\omega_q}{\omega_q^2}\right|\ll1,
\qquad
\left|\frac{d}{dt}\left(\frac{F_{qh}}{\omega_q}\right)\right|
\ll |F_{qh}|.
\label{eq:IV_adiabatic_criterion}
\end{equation}
These relative conditions apply away from zeros of
the source. At turning points, the absolute remainder
in Eq.~\eqref{eq:IV_exact_parts} and the boundary
terms must instead be controlled. When the source
varies on Hubble or Majoron-mass time scales,
$\max(H,m_\phi)/|\Delta E_q|\ll1$ provides a useful
modewise scale-separation test. It is not sufficient
without checking the actual source variation and the
initial transient. Uniform validity in exponentially
suppressed momentum tails is not required.

The leading local imaginary coherence in
Eq.~\eqref{eq:IV_ad_expansion}, together with the
current normalization
\eqref{eq:III_J_density_matrix}, gives
\begin{equation}
\langle J_{\rm PD}^0\rangle_{+,\rm loc}^{\rm ad}
=\frac{\chi_{\rm PD}(t)}{f_{\rm eff}}\dot\phi(t).
\label{eq:IV_Jlocal}
\end{equation}
Here the subscript ${\rm loc}$ identifies the
local particular-solution contribution; higher
local-derivative terms and finite-time boundary
contributions are not included. The reactive
susceptibility is
\begin{equation}
\chi_{\rm PD}(t)\equiv
\frac{2}{a^3(t)}\sum_h
\int\!\frac{d^3q}{(2\pi)^3}\,
\frac{A_q^2(t)D_{qh,i}}{\Delta E_q(t)}.
\label{eq:IV_chi}
\end{equation}
It has units of energy squared and is nonnegative
for the prepared ensemble. It is a reactive
coefficient, not a collision-induced damping rate.

For the collisionless benchmark, the name
$\chi_{\rm PD}^{\rm free}$ refers to this same
susceptibility with the initial occupations held fixed.
Differentiating Eq.~\eqref{eq:IV_chi} gives
\begin{equation}
\begin{aligned}
\chi_{\rm PD}^{\rm free}(t)&\equiv\chi_{\rm PD}(t),\\
\dot\chi_{\rm PD}^{\rm free}+3H\chi_{\rm PD}^{\rm free}
&=\frac{2}{a^3}\sum_h\int\!\frac{d^3q}{(2\pi)^3}\\
&\quad\times D_{qh,i}\frac{d}{dt}
\left(\frac{A_q^2}{\Delta E_q}\right).
\end{aligned}
\label{eq:IV_chi_free}
\end{equation}
This expression uses the occupations prepared in
Eq.~\eqref{eq:initial_FD_framework}; it does not
replace them by instantaneous equilibrium values.

Taking the divergence of the local current in
Eq.~\eqref{eq:IV_Jlocal} gives the current-based
local approximation to the microscopic transfer:
\begin{equation}
Q_{\rm micro}^{\rm ad}
=\frac{\dot\phi}{f_{\rm eff}^2}
\left[\chi_{\rm PD}\ddot\phi
 +(\dot\chi_{\rm PD}+3H\chi_{\rm PD})\dot\phi\right].
\label{eq:IV_Qlocal}
\end{equation}
Here $Q_{\rm micro}^{\rm ad}$ denotes the projected-
current, local adiabatic estimate; it does not denote
an exact evaluation of the full Ward source.
Inserting the same local current into the Majoron
equation \eqref{eq:phi_eom_framework} yields
\begin{equation}
\begin{aligned}
&\left(1+\frac{\chi_{\rm PD}}{f_{\rm eff}^2}\right)
\ddot\phi\\
&+\left[3H\left(1+\frac{\chi_{\rm PD}}{f_{\rm eff}^2}\right)
+\frac{\dot\chi_{\rm PD}}{f_{\rm eff}^2}\right]\dot\phi
+V_{,\phi}=0.
\end{aligned}
\label{eq:IV_phieom_local}
\end{equation}
With $Z_\phi(t)\equiv1+\chi_{\rm PD}(t)/f_{\rm eff}^2$,
this can be written as
\[
\frac{1}{a^3}\frac{d}{dt}
\left[a^3Z_\phi(t)\dot\phi\right]+V_{,\phi}=0.
\]
The modified inertia and time-dependent velocity
coefficient arise from the reactive local response.
The coefficient multiplying $\dot\phi$ is not,
by itself, a positive-definite irreversible
friction coefficient.

The distinction from the finite-time equation is
already visible at preparation. The exact initial
conditions imply $r_{qh}(t_i)=0$ and
$\dot\phi(t_i)=0$. Equation~\eqref{eq:III_scalar_density_matrix}
then gives a vanishing initial Ward source, and
Eq.~\eqref{eq:phi_eom_framework} gives
$\ddot\phi(t_i)=-V_{,\phi}(\phi_i)$.
In contrast, applying the local equation
\eqref{eq:IV_phieom_local} directly at $t_i$ would
give
\[
\ddot\phi_{\rm ad}(t_i)
=-\frac{V_{,\phi}(\phi_i)}
{1+\chi_{\rm PD}(t_i)/f_{\rm eff}^2}.
\]
The difference does not arise from a different
initial state. It arises because the local particular
solution omits the boundary response that enforces
the initial condition. The local approximation must
therefore not be imposed at $t_i$ without matching
to the full finite-time solution.

As a separate Ward consistency check, the real
part of the first local derivative correction is
\begin{equation}
\operatorname{Re}r_{qh}^{\rm ad}
=\frac{D_{qh,i}}{f_{\rm eff}\Delta E_q}
\frac{d}{dt}\left(\frac{A_q\dot\phi}{\Delta E_q}\right)
+O(\partial_t^3).
\label{eq:IV_real_adiabatic}
\end{equation}
Inserting this term into the projected scalar
bilinear and using the fixed-time identity
\eqref{eq:III_AB_Ward} reproduces the corresponding
current-divergence contribution except for the
geometric residual proportional to $\dot A_q$.
That residual was identified in
Eqs.~\eqref{eq:III_energy_continuity} and
\eqref{eq:IV_current_divergence}. The full Ward
identity remains exact, but the positive-energy
truncation alone does not demonstrate the cancellation
of the omitted basis contributions. Neither this
residual nor the local reactive term should be
interpreted as an independent decoherence rate.

\subsection{Nearly conserved Dirac limit and validity checks}
\label{subsec:IV_dirac_limit}

The formally large static factor $D_{qh}/\Delta E_q$ must not be interpreted as a finite physical current response in the exactly Dirac-symmetric theory. As $\delta m\to0$, the initially thermal population difference obeys
$D_{qh}=O(\Delta E_q(t_i))=O(\delta m)$.
At a fixed finite observation time, the exact retarded integral
\eqref{eq:IV_retarded_solution} has a bounded integration interval.
Expanding its sine and cosine kernels for small $\Delta E_q$ gives
\begin{equation}
\begin{aligned}
\operatorname{Re}r_{qh}(t)&=O(\delta m),\\
\operatorname{Im}r_{qh}(t)&=O((\delta m)^2),\\
\langle J_{\rm PD}^0(t)\rangle_+&=O((\delta m)^2),\\
Q_{\rm micro}^{(+),W}(t)&=O((\delta m)^2),
\end{aligned}
\label{eq:IV_dirac_finite_time}
\end{equation}
for fixed, sufficiently regular background evolution and momentum.
In the exact $\delta m=0$ theory, the prepared density matrix is proportional to the flavor identity and the conserved current cannot be sourced by this initial state. By contrast, the local zero-frequency adiabatic
expansion at fixed nonzero mass splitting yields
a finite ratio $D_{qh}/\Delta E_q$ as
$\delta m\to0$.

This finite ratio characterizes the formal
adiabatic particular solution, rather than
the dynamically induced current in the
finite-time collisionless initial-value problem.

The local approximation requires sufficiently
slow background evolution relative to
$|\Delta E_q|$ and control of the coherent
boundary transient. Since
$\Delta E_q=O(\delta m)$, the corresponding
microscopic time scale diverges as
$\delta m\to0$.

Furthermore, in the collisionless system the
oscillatory boundary transient does not
generally decay pointwise. Consequently,
the zero-frequency adiabatic approximation
is not uniform in the Dirac limit and cannot
be interchanged with the fixed-finite-time
limit of the exact retarded solution.

The validity of the leading treatment should be
assessed at three distinct levels. First, the instantaneous positive-frequency
reduction requires that the expansion-induced
positive-negative-energy mixing be suppressed.
A sufficient condition for occupied modes is
$H/E_{\alpha,q}\ll1$, as discussed in
Sec.~\ref{subsec:ward_time_basis}.
Possible positive-negative-energy mixing induced
by the Majoron interaction must also remain
perturbative within the reduced description. Second, the linear-response approximation
requires that the Majoron-induced mixing produce
only small changes in the initial occupation
difference. For a nonresonant, slowly varying
drive, a useful sufficient condition is
\begin{equation}
\epsilon_{\rm mix,q}(t)
\equiv
\frac{|A_q(t)\dot\phi(t)|}
{f_{\rm eff}|\Delta E_q(t)|}
\ll 1.
\label{eq:IV_weak_mixing}
\end{equation}
This condition is not necessary for a finite-time
initial-value problem, particularly near the
Dirac-symmetric limit.

More generally, the collisionless two-state
equations and the population invariant in
Eq.~\eqref{eq:IV_bloch_invariant} imply
\begin{equation*}
\frac{|D_{qh}(t)-D_{qh}(t_i)|}
{|D_{qh}(t_i)|}
\leq 2\mathcal G_q^2(t),
\end{equation*}
for nonzero initial occupation difference, where
\begin{equation*}
\mathcal G_q(t)
\equiv
\int_{t_i}^{t}dt'\,
\frac{|A_q(t')\dot\phi(t')|}
{f_{\rm eff}}.
\end{equation*}
Thus $\mathcal G_q\ll1$ provides an independent,
conservative sufficient condition for small
population changes over a finite interval.
Alternatively, the exact reduced two-state
evolution can be used to test the linearization
directly. Third, the local susceptibility approximation
requires additional adiabatic conditions,
including those stated in
Eq.~\eqref{eq:IV_adiabatic_criterion},
and control of the finite-time coherent boundary
transient. These additional restrictions are
not required for evaluating the finite-time
retarded solution in
Eq.~\eqref{eq:IV_retarded_solution}.
That solution remains exact only within the
linearized positive-frequency two-state system,
whose own validity conditions must still be
satisfied.

Finally, setting the collision term to zero is a physical limiting assumption, not a consequence of merely making a hidden scalar coupling flavor blind. If a mediator maintains appreciable scattering or pair-production rates after preparation, its coherent gain/loss terms must be retained in an interacting quantum kinetic equation
\cite{KainulainenParkkinen2024Collision}.
Here we intentionally isolate the collisionless problem in order to establish a clean baseline. Section~\ref{sec:cosmological_backreaction}
uses the susceptibility and, where necessary, the finite-time memory integral obtained here to quantify the resulting FRW Majoron backreaction, with no phenomenological dissipation parameter.

\section{Cosmological Majoron Backreaction}
\label{sec:cosmological_backreaction}

We now quantify the cosmological backreaction of the
coherent current derived in
Sec.~\ref{sec:coherent_kinetics}.
We use the prepared, subsequently collisionless
ensemble specified in
Sec.~\ref{subsec:ensemble_scope}
and the benchmark parameters of
Sec.~\ref{subsec:benchmark_scope}.
The Majoron evolves in a prescribed FRW background,
with its backreaction evaluated perturbatively.
We do not solve a fully self-consistent
Einstein--Majoron--fermion system.

The numerical results below evaluate the leading
local adiabatic response derived in
Sec.~\ref{subsec:IV_adiabatic}
and its effect on the Majoron evolution.
This response is the slowly varying local
particular-solution contribution to the finite-time
retarded kernel. It is not obtained by introducing
a phenomenological friction coefficient.

The exact finite-time solution also contains coherent
initial-boundary transients. We do not explicitly
resolve and average their complete oscillatory
evolution in the numerical results presented here.
The local contribution can therefore be interpreted
as a coarse-grained secular response only when
the relevant averaging or suppression of the
transients is independently justified.

Accordingly, the numerical results do not constitute
an exact evaluation of the complete instantaneous
memory backreaction or a general bound on the
omitted transient contributions. They also do not
include collision-induced effects.
These distinctions are retained when assessing
the cosmological significance of the response.

\subsection{Background expansion and benchmark initial conditions}
\label{subsec:V_background}

We prescribe a spatially flat $\Lambda$CDM expansion,
\begin{equation}
\begin{aligned}
H^2(a)&=H_0^2\left[\Omega_r a^{-4}+\Omega_m a^{-3}
+\Omega_\Lambda\right],\\
\Omega_\Lambda&=1-\Omega_m-\Omega_r.
\end{aligned}
\label{eq:V_H}
\end{equation}
Here $H_0$ is the present Hubble parameter, while
$\Omega_r$, $\Omega_m$, and $\Omega_\Lambda$ are the
present radiation, matter, and cosmological-constant
density fractions. We use the fixed inputs
\begin{equation}
\begin{aligned}
H_0&=1.44\times10^{-33}\,\mathrm{eV},\\
\Omega_m&=0.315,\qquad \Omega_r=9.2\times10^{-5},\\
\Omega_\Lambda&\simeq0.685.
\end{aligned}
\label{eq:V_cosmology}
\end{equation}
These values are representative of the Planck-era
$\Lambda$CDM background \cite{Planck:2018vyg}.
They are prescribed inputs, not parameters inferred
from the hidden-sector dynamics.

The Majoron potential was defined in
Eq.~\eqref{eq:majoron_potential}. For the numerical
benchmark, we use the following scalar scales and
the corresponding potential derivative:
\begin{equation}
\begin{aligned}
m_\phi&=10^{-33}\,\mathrm{eV},\qquad
f_\phi=10^{9}\,\mathrm{GeV},\\
V_{,\phi}(\phi)&=m_\phi^2f_\phi\sin(\phi/f_\phi).
\end{aligned}
\label{eq:V_V}
\end{equation}
The fermion masses and derivative-coupling scale
are those in Eq.~\eqref{eq:benchmark_framework}.
For the preparation epoch and temperature, we take
\begin{equation}
\begin{aligned}
a_i&=10^{-2},\qquad
T_{\nu0}=1.68\times10^{-4}\,\mathrm{eV},\\
T_{\rm PD0}&=0.4T_{\nu0}
\simeq6.70\times10^{-5}\,\mathrm{eV},\\
T_{\rm PD,i}&=\frac{T_{\rm PD0}}{a_i}
\simeq6.70\times10^{-3}\,\mathrm{eV}.
\end{aligned}
\label{eq:V_benchmark}
\end{equation}
Here $T_{\nu0}$ is the present reference neutrino
temperature, and $T_{\rm PD0}$ is a convenient
comoving momentum scale for the prepared ensemble.
It does not represent an equilibrium temperature
maintained by the collisionless fermions at $a=1$.
The choices of $a_i$ and $T_{\rm PD,i}$ specify an
idealized thermal preparation, not a decoupling epoch
obtained from calculated interaction rates.

The initial occupation difference in
Eq.~\eqref{eq:IV_initial} remains fixed at comoving
momentum after preparation. Under the same comoving
temperature normalization and relativistic preparation,
earlier decoupling would give a smaller difference:
in the small-splitting regime it scales approximately
as the square of the preparation scale factor.
Free streaming cannot regenerate the occupation
difference associated with a later equilibrium state.
This observation depends on the specified preparation
history and is not a universal statement about all
earlier-decoupling scenarios.

Using the e-fold variable introduced in
Sec.~\ref{subsec:IV_exact_solution}, let
$\theta\equiv\phi/f_\phi$. We impose
\begin{equation}
\theta(\mathcal N_i)=1,\qquad
\theta'(\mathcal N_i)=0,\qquad
\mathcal N_i=\ln(10^{-2}),
\label{eq:V_initial_phi}
\end{equation}
where a prime denotes differentiation with respect
to $\mathcal N$. The fermion density matrix retains
the preparation conditions in
Eq.~\eqref{eq:initial_rho_framework}.
For the stated Majoron initial condition,
$\rho_\phi(t_i)=m_\phi^2f_\phi^2(1-\cos1)
\simeq4.60\times10^{-31}\,\mathrm{eV}^4$.
This energy density is negligible compared with the
background cosmological components, consistently with
the spectator assumption in
Sec.~\ref{subsec:benchmark_scope}.

The Majoron contribution to $H(a)$ is neglected.
The hidden-fermion contribution is likewise not
separately included in the prescribed expansion.
The calculated backreaction therefore measures the
portal-induced change relative to the uncoupled
Majoron evolution in the same external background;
it is not a calculation of the total gravitational
effect of the hidden ensemble.

\subsection{Free-streaming susceptibility and redshift scaling}
\label{subsec:V_susceptibility}

The local reactive susceptibility was defined in
Eq.~\eqref{eq:IV_chi_free}. The prepared occupation
difference $D_{qh,i}$ is held fixed at comoving momentum.
Its subsequent scale-factor dependence comes only from
physical dilution, the spinor overlap, and the evolving
energy splitting. For numerical integration, use the
dimensionless comoving momentum $x\equiv q/T_{\rm PD0}$,
with $T_{\rm PD0}$ specified in Eq.~\eqref{eq:V_benchmark}.
The helicity-independent preparation and overlap give
\begin{equation}
\chi_{\rm PD}^{\rm free}(a)
=\frac{2T_{\rm PD0}^3}{\pi^2a^3}
\int_0^\infty dx\,x^2
\left.\frac{A_q^2(a)D_{qh,i}}{\Delta E_q(a)}
\right|_{q=xT_{\rm PD0}}.
\label{eq:V_chi_free}
\end{equation}
The explicit helicity sum in Eq.~\eqref{eq:IV_chi}
has been performed; there is no additional spin factor.

The small-splitting expansion of the prepared occupation
difference was given in Eq.~\eqref{eq:occupation}.
For numerical evaluation, the equivalent exact expression
\begin{equation}
D_{qh,i}=
\frac{\sinh\!\left[\frac{\Delta E_q(a_i)}{2T_{\rm PD,i}}\right]}
{\cosh\!\left[\frac{E_q(t_i)}{T_{\rm PD,i}}\right]
 +\cosh\!\left[\frac{\Delta E_q(a_i)}{2T_{\rm PD,i}}\right]}
\label{eq:V_Dinit}
\end{equation}
avoids subtracting nearly equal Fermi--Dirac occupations.
Similarly, the energy splitting is best evaluated as
\begin{equation}
\Delta E_q(a)=
\frac{\delta m(m_1+m_2)}{E_{1,q}(a)+E_{2,q}(a)},
\label{eq:V_stable_delta}
\end{equation}
which follows exactly by rationalizing
$E_{1,q}-E_{2,q}$. At fixed comoving momentum,
\begin{equation}
\frac{d\ln\Delta E_q}{d\mathcal N}
=\frac{p^2}{E_{1,q}E_{2,q}}.
\label{eq:V_omega_derivative}
\end{equation}
The splitting therefore grows approximately as $a$
for relativistic modes and approaches $\delta m$ for
nonrelativistic modes.

Define the dimensionless local response
$\epsilon(a)\equiv\chi_{\rm PD}^{\rm free}(a)/f_{\rm eff}^2$
and $\epsilon_{\mathcal N}\equiv d\epsilon/d\mathcal N$.
Because $D_{qh,i}$ is time independent, differentiating
the susceptibility gives
\begin{equation}
\begin{aligned}
\epsilon_{\mathcal N}+3\epsilon
&=\frac{2}{a^3f_{\rm eff}^2}
\sum_h\int\!\frac{d^3q}{(2\pi)^3}\,
D_{qh,i}\\
&\quad\times\frac{d}{d\mathcal N}
\left(\frac{A_q^2}{\Delta E_q}\right).
\end{aligned}
\label{eq:V_epsilon_derivative}
\end{equation}
For nearly degenerate masses, $A_q\simeq1$.
The derivative on the right is then negative at
nonzero momentum, since $\Delta E_q$ grows as
$p=q/a$ decreases.

For modes that remain relativistic, $p\gg m_D$,
we have $\Delta E_q\simeq m_D\delta m/p\propto a$.
Using the prepared distribution and the explicit
helicity sum, the leading result is
\begin{equation}
\begin{aligned}
\chi_{\rm PD}^{\rm free}(a)
&\simeq\frac{T_{\rm PD,i}^2}{3}
\left(\frac{a_i}{a}\right)^4,\\
\frac{\epsilon_{\mathcal N}}{\epsilon}&\simeq-4.
\end{aligned}
\label{eq:V_a4}
\end{equation}
The normalization follows from
$\int_0^\infty dx\,x^2n_F(x)[1-n_F(x)]=\pi^2/6$.
The prefactor $1/3$ includes both helicities and
the current-trace normalization.
For the benchmark in Eq.~\eqref{eq:V_benchmark},
$T_{\rm PD,i}^2/3=1.50\times10^{-5}\,\mathrm{eV}^2$,
consistent with the initial numerical susceptibility.
Thus $\epsilon_{\mathcal N}+3\epsilon\simeq-\epsilon$
in the relativistic regime.

The $a^{-4}$ behavior applies only while the
momentum modes dominating the integral are
relativistic. When these modes become deeply
nonrelativistic, $\Delta E_q\to\delta m$ and
$A_q\to1$; the frozen occupations then give
$\chi_{\rm PD}^{\rm free}\propto a^{-3}$.

Using $A_q^2\leq 1$, the monotonic growth
of $\Delta E_q(a)$, and the initial
Fermi--Dirac occupations, we obtain
an upper bound on the susceptibility.
Writing $T_i\equiv T_{{\rm PD},i}$ and
$p_i\equiv q/a_i$, the mean value theorem gives
\begin{eqnarray}
\frac{D_{qh,i}}{\Delta E_q(a_i)}
&=
\frac{n_F(\xi_q/T_i)
      [1-n_F(\xi_q/T_i)]}{T_i} \nonumber
\\
&\leq
\frac{n_F(p_i/T_i)
      [1-n_F(p_i/T_i)]}{T_i}.\nonumber
\end{eqnarray}
where $\xi_q$ lies between
$E_{1,q}(t_i)$ and $E_{2,q}(t_i)$.
Here we have used $E_{\alpha,q}(t_i)\geq p_i$
and the monotonic decrease of
$n_F(x)[1-n_F(x)]$ for $x\geq 0$.
Consequently,
\begin{equation}
\begin{aligned}
\epsilon(a)
&\leq
\frac{2T_i^2}{\pi^2 f_{\rm eff}^2}
\left(\frac{a_i}{a}\right)^3
\int_0^\infty dx\,x^2 n_F(x)[1-n_F(x)]
\\
&=
\frac{T_i^2}{3f_{\rm eff}^2}
\left(\frac{a_i}{a}\right)^3.
\end{aligned}
\end{equation}
Thus,
\begin{equation}
0\leq\epsilon(a)\leq
\frac{T_{{\rm PD},i}^2}{3f_{\rm eff}^2}
\left(\frac{a_i}{a}\right)^3,
\qquad a\geq a_i.
\label{eq:V_a3}
\end{equation}

Unlike the relativistic scaling in Eq.~\ref{eq:V_a4}, the upper bound in Eq.~\ref{eq:V_a3} remains valid in the nonrelativistic regime.
For our benchmark, $T_{\rm PD,i}/f_{\rm eff}=3.30\times10^{-23}$, yielding $\epsilon(a)\leq3.64\times10^{-46}(a_i/a)^3$ and hence $\epsilon(1)\leq3.64\times10^{-52}$. Thus, the local reactive susceptibility correction is extremely small throughout the considered evolution, independently of the detailed momentum integration.

For comparison, consider a hypothetical system
that instead maintains zero-chemical-potential
Fermi--Dirac occupations at
$T_{\rm PD}(a)=T_{\rm PD,i}a_i/a$.
Its relativistic static susceptibility scales as
$T_{\rm PD}^2\propto a^{-2}$, rather than $a^{-4}$.
While the relevant modes remain relativistic, the
ratio of this hypothetical equilibrium susceptibility
to the collisionless result grows approximately as
$(a/a_i)^2$. For the specified benchmark, direct
momentum integration at $a=1$ gives a ratio of about
$9.93\times10^3$, close to the relativistic estimate
$10^4$. This comparison illustrates the different
preparation and evolution assumptions; the equilibrium
system is not part of our collisionless benchmark.
In particular, replacing the fixed $D_{qh,i}$ by
instantaneous thermal occupations would significantly
overestimate its late-time local susceptibility.

These scalings characterize the local reactive
coefficient. They do not by themselves bound the
complete finite-time boundary transient discussed in
Sec.~\ref{subsec:IV_adiabatic}.

\subsection{Effective FRW evolution from the coherent response}
\label{subsec:V_friedmann_evolution}

The local current and its current-based energy-transfer
rate were derived in Eqs.~\eqref{eq:IV_Jlocal} and
\eqref{eq:IV_Qlocal}. We now express their cosmological
effect in the e-fold variable used in
Sec.~\ref{subsec:V_background}. Define
$h_{\mathcal N}\equiv d\ln H/d\mathcal N$.
With the dimensionless response $\epsilon$ from
Sec.~\ref{subsec:V_susceptibility}, the local transfer
rate can be written in a form suitable for numerical
evaluation:
\begin{equation}
\begin{aligned}
Q_{\rm micro}^{\rm ad}
={}&f_\phi^2H^3\bigl\{
\epsilon\,\theta'\bigl(\theta''+h_{\mathcal N}\theta'\bigr)\\
&\hspace{19mm}+(\epsilon_{\mathcal N}+3\epsilon)(\theta')^2
\bigr\}.
\end{aligned}
\label{eq:V_Q}
\end{equation}
Equation~\eqref{eq:V_Q} is the e-fold form of
Eq.~\eqref{eq:IV_Qlocal}. Its sign may change during
cosmological evolution; it is not an assumed
positive-definite friction law.

The local scalar equation itself was already given
in Eq.~\eqref{eq:IV_phieom_local}. To convert it to
e-fold time, we use the kinematic relations
\begin{equation}
\begin{aligned}
\dot\phi&=f_\phi H\theta',\\
\ddot\phi&=f_\phi H^2
\bigl(\theta''+h_{\mathcal N}\theta'\bigr).
\end{aligned}
\label{eq:V_local_eom}
\end{equation}
The resulting local equation is
\begin{equation}
\begin{aligned}
&(1+\epsilon)
\left[\theta''+(3+h_{\mathcal N})\theta'\right]
+\epsilon_{\mathcal N}\theta'\\
&\hspace{30mm}+\frac{m_\phi^2}{H^2}\sin\theta=0.
\end{aligned}
\label{eq:V_eom_efold}
\end{equation}
For the prescribed background in
Eq.~\eqref{eq:V_H}, the Hubble derivative is
\begin{equation}
h_{\mathcal N}=-
\frac{2\Omega_r a^{-4}+\tfrac32\Omega_m a^{-3}}
{\Omega_r a^{-4}+\Omega_m a^{-3}+\Omega_\Lambda}.
\label{eq:V_hprime}
\end{equation}

Because $\epsilon\ll1$, subtracting two independently
integrated full scalar solutions would lose numerical
precision. Instead write
$\theta=\theta_0+\delta\theta+O(\epsilon^2)$,
where $\theta_0$ is the uncoupled solution and
$\delta\theta$ is the first-order portal-induced change.
The uncoupled equation is
\begin{equation}
\theta_0''+(3+h_{\mathcal N})\theta_0'
+\frac{m_\phi^2}{H^2}\sin\theta_0=0.
\label{eq:V_background_theta}
\end{equation}
Substituting the perturbative decomposition into
Eq.~\eqref{eq:V_eom_efold} and retaining first order
gives
\begin{equation}
\begin{aligned}
&\delta\theta''+(3+h_{\mathcal N})\delta\theta'
+\frac{m_\phi^2}{H^2}\cos\theta_0\,\delta\theta\\
&\qquad=\epsilon\frac{m_\phi^2}{H^2}\sin\theta_0
-\epsilon_{\mathcal N}\theta_0'.
\end{aligned}
\label{eq:V_delta_eom}
\end{equation}
We impose $\delta\theta(\mathcal N_i)=
\delta\theta'(\mathcal N_i)=0$, consistent with
identical initial field values and velocities.
The fractional velocity change is
\begin{equation}
\frac{\delta\dot\phi}{\dot\phi_0}
=\frac{\delta\theta'}{\theta_0'},
\qquad \theta_0'\neq0.
\label{eq:V_velocity_change}
\end{equation}
At this order, $Q_{\rm micro}^{\rm ad}$ in
Eq.~\eqref{eq:V_Q} can be evaluated using $\theta_0$
in place of $\theta$, with corrections of second
order in the local response.

For numerical stability, we integrate the perturbation
as $\delta\theta/10^{-52}$ and
$\delta\theta'/10^{-52}$. This rescales the tiny
response without changing the differential equation.
For the chosen benchmark, the resulting perturbation
is of order $10^{-52}$. We check convergence under
changes of the rescaling and solver tolerances.
Subtracting the full solutions would not resolve such
a small change in ordinary double precision.

The local evolution assumes the adiabatic conditions
and boundary-transient control discussed in
Sec.~\ref{subsec:IV_adiabatic}. Although the chosen
initial values of $\delta\theta$ and $\delta\theta'$ match
the finite-time preparation conditions, the local equation
need not match the exact initial acceleration because its
particular solution omits the initial boundary response.
The result here therefore characterizes the local
reactive backreaction, not the full finite-time
nonlocal evolution.

\subsection{Numerical benchmark results}
\label{subsec:V_results}

\begin{figure}[!b]
\centering
\includegraphics[width=\columnwidth]{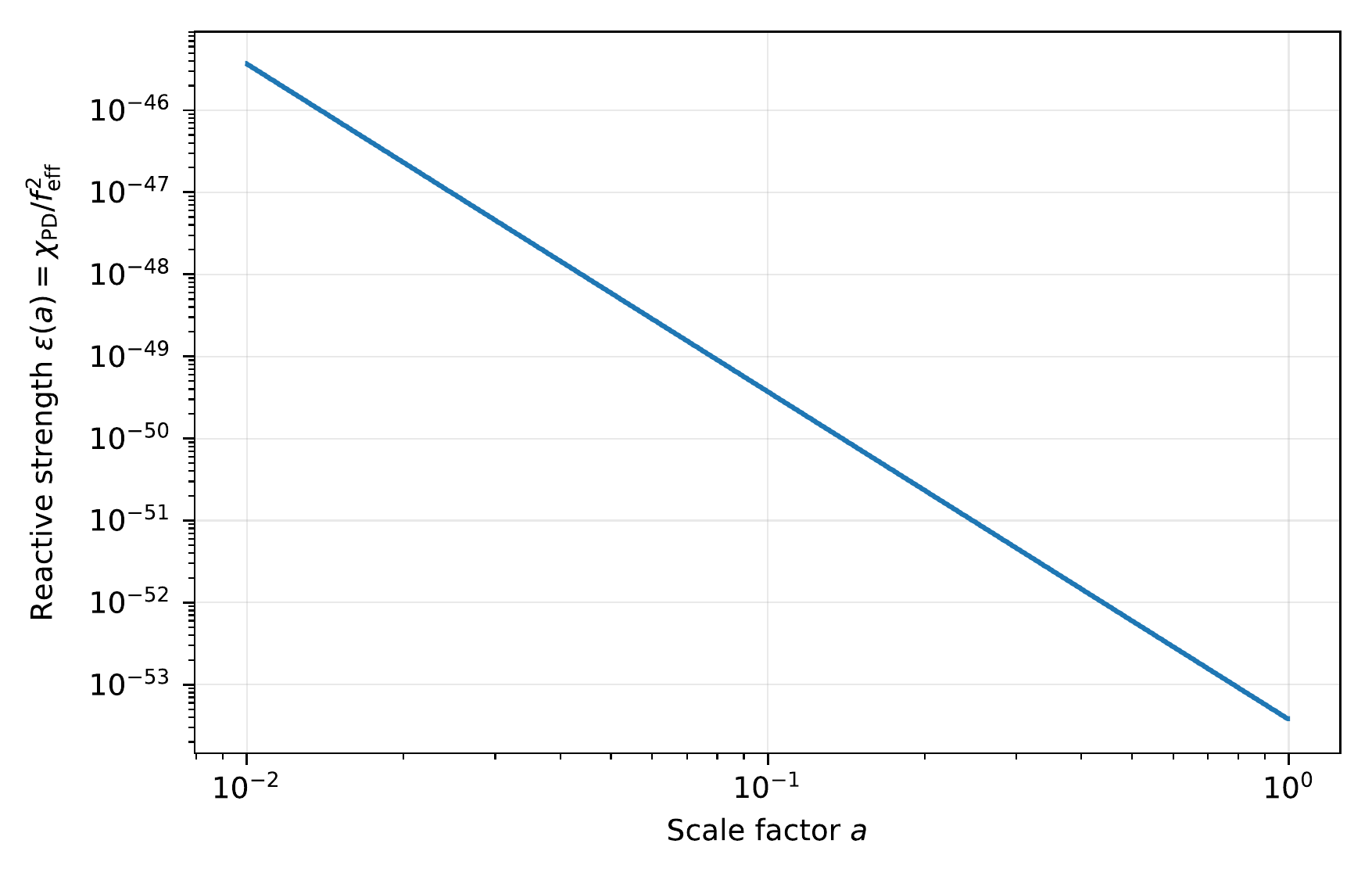}
\caption{Dimensionless local reactive response
$\epsilon(a)=\chi_{\rm PD}^{\rm free}(a)/f_{\rm eff}^2$
for the prepared, subsequently collisionless ensemble.
The approximate $a^{-4}$ behavior in the relativistic regime
follows from fixed comoving occupations, not ongoing
thermal equilibrium.}
\label{fig:V_epsilon}
\end{figure}

We evaluate the comoving-momentum integrals using the variable
$x$ introduced in Sec.~\ref{subsec:V_susceptibility}, with
260-point Gauss--Legendre quadrature on $0\leq x\leq48$.
The uncoupled Majoron equation and the linear perturbation
Eq.~\eqref{eq:V_delta_eom} are integrated from $a_i$ to $a=1$ with an adaptive high-order Runge--Kutta method.
The initial fermion occupations are held fixed at comoving
momentum, as specified in Sec.~\ref{subsec:ensemble_scope}.
We denote the redshift by $z$, with $a=(1+z)^{-1}$.
Figure~\ref{fig:V_epsilon} and Table~\ref{tab:V_chi}
show the resulting local reactive response.

\begin{table}[!b]
\centering
\caption{Free-streaming susceptibility and dimensionless local  response for the benchmark of
Eqs.~\eqref{eq:benchmark_framework},
\eqref{eq:V_cosmology}, and \eqref{eq:V_benchmark}.
The entries describe the local adiabatic coefficient,
not the complete finite-time response.}
\label{tab:V_chi}
\begin{tabular}{rcc}
\hline\hline
$z$ & $\chi_{\rm PD}^{\rm free}$ [$\mathrm{eV}^2$]
& $\epsilon$ \\
\hline
99 & $1.50\times10^{-5}$  & $3.64\times10^{-46}$ \\
9  & $1.50\times10^{-9}$  & $3.64\times10^{-50}$ \\
2  & $1.21\times10^{-11}$ & $2.95\times10^{-52}$ \\
1  & $2.40\times10^{-12}$ & $5.82\times10^{-53}$ \\
0  & $1.50\times10^{-13}$ & $3.65\times10^{-54}$ \\
\hline\hline
\end{tabular}
\end{table}

For the uncoupled Majoron, the present epoch
($\mathcal N=0$, or $z=0$) gives
\begin{equation}
\begin{aligned}
\theta_0(0)&=0.944,\\
\dot\phi_0(0)&=-1.53\times10^{-16}\,\mathrm{eV}^2,\\
\rho_\phi(0)&=4.25\times10^{-31}\,\mathrm{eV}^4.
\end{aligned}
\label{eq:V_phi_today}
\end{equation}
The scalar energy density is approximately
$1.69\times10^{-20}$ of the vacuum-energy density
in the prescribed background, consistent with the
spectator treatment.

Table~\ref{tab:V_Q} gives the leading local adiabatic
energy-transfer rate and the integrated fractional
velocity perturbation at selected redshifts.
At $z=0$, we find
\begin{equation}
\begin{aligned}
Q_{\rm micro}^{\rm ad}(0)
&=-3.85\times10^{-119}\,\mathrm{eV}^5,\\
\frac{Q_{\rm micro}^{\rm ad}(0)}{3H_0\rho_\phi(0)}
&=-2.10\times10^{-56},\\
\frac{\delta\dot\phi(0)}{\dot\phi_0(0)}
&=+5.23\times10^{-53}.
\end{aligned}
\label{eq:V_main_results}
\end{equation}
Because $\dot\phi_0(0)<0$, the positive fractional
velocity shift corresponds to a minute increase in
$|\dot\phi|$, not to freezing. This conclusion is
restricted to the computed local perturbative response.

\begin{table}[t]
\centering
\caption{Leading local adiabatic energy exchange and the
accumulated fractional Majoron velocity perturbation.
These results do not resolve the undamped finite-time
boundary transient.}
\label{tab:V_Q}
\begin{tabular}{rcc}
\hline\hline
$z$ & $Q_{\rm micro}^{\rm ad}$ [$\mathrm{eV}^5$]
& $\delta\dot\phi/\dot\phi_0$ \\
\hline
9   & $+2.49\times10^{-116}$ & $-3.63\times10^{-50}$ \\
2   & $+1.04\times10^{-117}$ & $-2.43\times10^{-52}$ \\
1   & $+2.47\times10^{-118}$ & $-5.33\times10^{-54}$ \\
0.5 & $+3.50\times10^{-119}$ & $+3.54\times10^{-53}$ \\
0   & $-3.85\times10^{-119}$ & $+5.23\times10^{-53}$ \\
\hline\hline
\end{tabular}
\end{table}

\begin{figure}[t]
\centering
\includegraphics[width=\columnwidth]{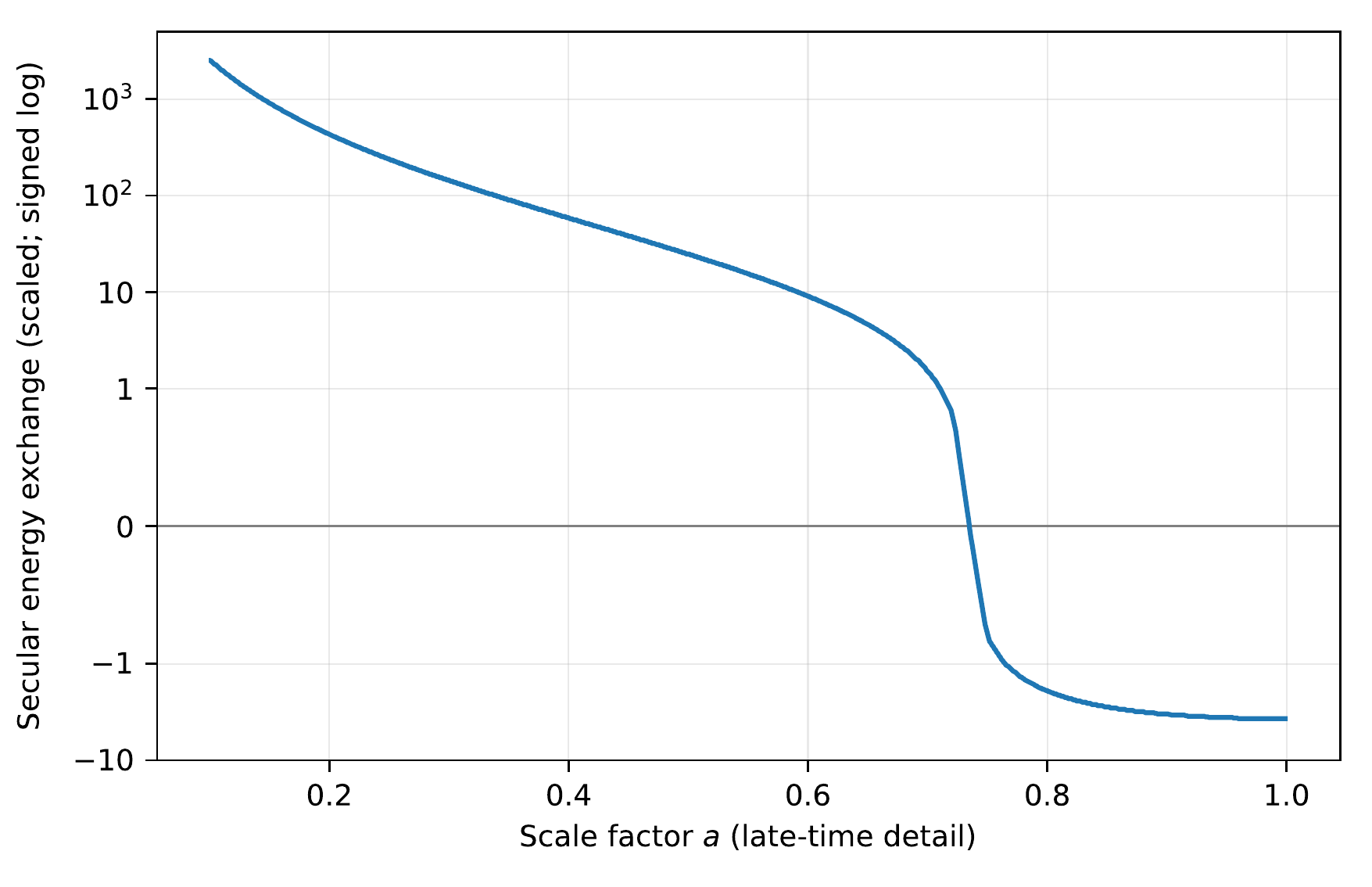}
\caption{Leading local adiabatic energy exchange
$Q_{\rm micro}^{\rm ad}(a)$ in units of
$10^{-119}\,\mathrm{eV}^5$, shown for
$0.1\leq a\leq1$ on a signed-logarithmic ordinate.
Its sign reversal demonstrates that this local response
is not a positive-definite friction law. The complete
finite-time oscillatory contribution is not plotted.}
\label{fig:V_Q}
\end{figure}

Figure~\ref{fig:V_Q} shows the continuous local
energy-transfer history. The entries in
Table~\ref{tab:V_Q} correspond to
$a=0.1,1/3,1/2,2/3$, and $1$, respectively.
The local transfer rate changes sign at $a\simeq0.735$
($z\simeq0.361$). The fractional velocity perturbation
changes sign earlier, at $a\simeq0.512$
($z\simeq0.953$). These crossings need not coincide:
$Q_{\rm micro}^{\rm ad}$ measures the local rate of
energy exchange, whereas the velocity perturbation
reflects the accumulated effect of the portal since
preparation. A positive local transfer rate therefore
does not necessarily imply a negative velocity shift
at the same epoch. Both sign changes are properties
of the computed reactive local response, not evidence
for a negative dissipation coefficient.

The two contributions to Eq.~\eqref{eq:V_Q} have
opposite signs at the present epoch:
\begin{equation}
\begin{aligned}
\epsilon\dot\phi_0\ddot\phi_0
&=+8.37\times10^{-119}\,\mathrm{eV}^5,\\
H(\epsilon_{\mathcal N}+3\epsilon)\dot\phi_0^2
&=-1.22\times10^{-118}\,\mathrm{eV}^5.
\end{aligned}
\label{eq:V_Q_parts}
\end{equation}
Their sum reproduces the negative present-day value
in Eq.~\eqref{eq:V_main_results}. In the relativistic
free-streaming regime,
$\epsilon_{\mathcal N}\simeq-4\epsilon$, so the
second term approaches $-H\epsilon\dot\phi^2$;
the first depends on the instantaneous scalar
acceleration. Their competition permits a reversal
of $Q_{\rm micro}^{\rm ad}$ without introducing
irreversible or negative friction.

These numerical results evaluate the slowly varying
local particular-solution contribution. No explicit
phase averaging of the complete retarded kernel has
been performed here. The finite-time boundary and memory contributions are quantitatively controlled within the linearized collisionless positive-energy two-state framework
in Sec.~\ref{subsec:V_checks}.
This comparison supports the accuracy of the
local adiabatic velocity response for the
specified benchmark without assuming
phenomenological damping.

\subsection{Adiabatic control, Ward consistency, and numerical errors}
\label{subsec:V_checks}

We first examine the separation of time scales
for modes that contribute appreciably to the
prepared hidden pseudo-Dirac ensemble.
For the representative comoving momentum
$q=3.15T_{\rm PD0}$, we obtain
\begin{equation}
\begin{array}{lcc}
 & a=0.01 & a=1 \\
\Delta E_q\ [\mathrm{eV}]
 & 9.47\times10^{-14} & 9.46\times10^{-12} \\
\Delta E_q/H
 & 1.16\times10^{17} & 6.58\times10^{21}.
\end{array}
\label{eq:V_hierarchy}
\end{equation}
Using the present-day uncoupled field velocity
in Eq.~\eqref{eq:V_phi_today}, we find
$\epsilon_{\rm mix,q}(a=1)\simeq
7.98\times10^{-26}$
for the same mode.
At preparation, this parameter vanishes
because $\dot\phi(t_i)=0$.
These values support the weak-driving and
scale-separation approximations for this mode.
However, a large $\Delta E_q/H$ does not
by itself establish the accuracy of the
local response.
The source-variation condition in
Eq.~\eqref{eq:IV_adiabatic_criterion}
must also be satisfied in the relevant momentum region.

A further issue arises from the finite
preparation time.
The exact retarded solution in
Eq.~\eqref{eq:IV_retarded_solution}
satisfies the initial condition
$\rho_{12,qh}(t_i)=0$.
Although the leading boundary term in
Eq.~\eqref{eq:IV_exact_parts}
vanishes because the initial driving
source is zero, the next boundary term
need not vanish.
It is proportional to the initial
scalar acceleration, which is generally
nonzero. This boundary response is required
to match the prepared quantum state.
It cannot be removed solely by assuming
$\Delta E_q/H\gg1$.

The numerical results in
Tables~\ref{tab:V_chi} and \ref{tab:V_Q}
and Fig.~\ref{fig:V_Q}
describe the local adiabatic contribution.
They do not result from an explicit
phase average of the complete retarded
current.
We therefore evaluate the difference
between the finite-time retarded response
and the local adiabatic response.
Both calculations use the same prepared
fermionic state, uncoupled Majoron
trajectory, and prescribed FRW background.

For each comoving mode, we use the
driving source $F_{qh}$ and coherence
frequency $\omega_q$ defined in
Eq.~\eqref{eq:IV_omega_s}.
The source is evaluated along the
uncoupled Majoron trajectory $\phi_0(t)$.
We also define
\begin{equation}
L_{qh}(t)=
\frac{1}{\omega_q(t)}
\frac{d}{dt}
\left[
\frac{F_{qh}(t)}{\omega_q(t)}
\right],
\label{eq:V_L_def}
\end{equation}
where $L_{qh}$ is the first derivative
correction to the leading local
coherence response.

Two integrations by parts give
\begin{equation}
\begin{aligned}
\rho_{12,qh}^{\rm ret}(t)
={}&-\frac{i F_{qh}(t)}{\omega_q(t)}
+L_{qh}(t)\\
&-L_{qh}(t_i)e^{-i\Theta_q(t,t_i)}\\
&-\int_{t_i}^{t}dt'\,
e^{-i\Theta_q(t,t')}\dot L_{qh}(t').
\end{aligned}
\label{eq:V_retarded_parts}
\end{equation}
The first term is the leading local
adiabatic coherence.
The third term is the initial boundary
transient, and the last term retains
the remaining temporal memory.
The boundary term is generally nonzero
even though $F_{qh}(t_i)=0$.

Because $L_{qh}$ is real, the exact
decomposition implies
\begin{equation}
\begin{aligned}
\left|
\operatorname{Im}\rho_{12,qh}^{\rm ret}
+\frac{F_{qh}}{\omega_q}
\right|
\leq{}&
|L_{qh}(t_i)|\\
&+\int_{t_i}^{t}dt'\,
|\dot L_{qh}(t')|.
\end{aligned}
\label{eq:V_modewise_bound}
\end{equation}
This modewise inequality bounds the
difference between the retarded and
leading local contributions to the
current. It explicitly retains the initial
boundary response and the remaining
memory term. No collision term, phenomenological
damping, or phase-averaging assumption is introduced.

We integrate this inequality over
the prepared comoving momentum
distribution and propagate the resulting
current-difference envelope through the
linearized Majoron equation.
We write
\begin{equation}
j(t)=
\frac{\langle J_{\rm PD}^0(t)\rangle_+}
{f_{\rm eff}f_\phi H(t)},
\qquad
|j_{\rm ret}-j_{\rm ad}|
\leq\beta,
\label{eq:V_j_beta}
\end{equation}
where $j$ is the dimensionless projected
current source and $\beta$ is its
nonnegative difference envelope.
The subscripts ``ret'' and ``ad''
denote the finite-time retarded and
leading local adiabatic responses,
respectively.

For the differences between the two
scalar responses, a nonnegative
comparison system propagates this
current bound into a bound on the
fractional velocity difference. Its derivation, initial conditions, analytical source
derivatives, and numerical implementation are given
in Appendix~\ref{app:nonlocal_comparison}.
At $a=1$, the numerical evaluation gives
\begin{equation}
\begin{aligned}
R_{\rm ad}
&=+5.23\times10^{-53},\\
|R_{\rm ret}-R_{\rm ad}|
&\lesssim3.54\times10^{-69},\\
\frac{|R_{\rm ret}-R_{\rm ad}|}{|R_{\rm ad}|}
&\lesssim6.78\times10^{-17}.
\end{aligned}
\label{eq:V_retarded_velocity_bound}
\end{equation}
where $R=\delta\dot\phi/\dot\phi_0$
is the fractional change in the
Majoron velocity relative to the
uncoupled solution.
The comparison includes both the
finite-time initial boundary transient
and the temporal memory contribution.

This result provides a quantitative
test of the local approximation at the
level of the cosmological velocity
response. In particular, it goes beyond the
instantaneous hierarchy
$\Delta E_q/H\gg1$ by controlling
the difference between the two
current-driven scalar responses.
The comparison uses an inequality
within the linearized, collisionless
positive-energy two-state system.
It is not obtained by subtracting two
independently integrated full scalar
trajectories. The numerical envelope is tested for
convergence, but its evaluation is not
interval-certified.

A separate consistency check concerns the
microscopic Ward identity and the
time-dependent positive-energy projection.
The fixed-time relation in
Eq.~\eqref{eq:III_AB_Ward}
makes the commutator contribution to
the projected current divergence agree
with the projected scalar bilinear.
However, the time dependence of the
spinor overlap produces the additional
$\dot A_q$ term in
Eq.~\eqref{eq:IV_current_divergence}.
This term arises from the evolving
positive-energy projection.
It is neither an independent
dissipative source nor a violation
of the exact operator Ward identity.
The geometric connection restores
the exact identity in the complete
spinor space, as shown in
Eq.~\eqref{eq:III_covariant_eigenbasis_Ward}.

For $q=3.15T_{\rm PD0}$ at $a=1$,
we obtain
\begin{equation}
1-A_q\simeq1.12\times10^{-13},
\qquad
\left|
\frac{d\ln A_q}{d\mathcal N}
\right|
\simeq2.22\times10^{-13}.
\label{eq:V_overlap_today}
\end{equation}
These values show that the overlap
and its evolution are close to their
Dirac-limit behavior for this mode.
They do not independently bound the
initial boundary transient or provide
a relative bound on the integrated
Ward residual near a zero of the
net energy-transfer rate.

The projected Ward-source discrepancy is evaluated
independently using the leading local coherence
in Eq.~\eqref{eq:IV_ad_expansion}.
From Eq.~\eqref{eq:III_energy_continuity}, the
difference between the projected Ward source and
the divergence of the projected current is
$Q_{\rm micro}^{(+),W}-Q_{\rm micro}^{(+),J}
=-H\dot\phi^2\epsilon_A$, where
\begin{equation}
\epsilon_A(\mathcal N)
=
\frac{2}{a^3f_{\rm eff}^2}
\sum_h\int\frac{d^3q}{(2\pi)^3}
\frac{A_q^2D_{qh,i}}{\Delta E_q}
\frac{d\ln A_q}{d\mathcal N}.
\label{eq:V_Ward_projection_epsilon}
\end{equation}
Here $\epsilon_A$ measures the contribution from
the time dependence of the positive-energy current
overlap. It is distinct from the susceptibility
ratio $\epsilon$ defined in
Sec.~\ref{subsec:V_susceptibility}.

Writing $\Delta\theta_W$ for the difference between
the Ward-source-driven and projected-current-driven
scalar perturbations, we obtain
\begin{equation}
\begin{aligned}
&\Delta\theta_W''
+(3+h_{\mathcal N})\Delta\theta_W'
+\frac{m_\phi^2}{H^2}
\cos\theta_0\,\Delta\theta_W\\
&\qquad=\epsilon_A\theta_0'.
\end{aligned}
\label{eq:V_Ward_projection_response}
\end{equation}
We impose
$\Delta\theta_W(\mathcal N_i)
=\Delta\theta_W'(\mathcal N_i)=0$.
The source is evaluated from the same prepared
fermion distribution and uncoupled Majoron
background used in the local calculation.

Independent momentum integration and perturbative
FRW evolution give
$\epsilon_A(0)\simeq-1.92\times10^{-66}$ and
$\Delta\theta_W'(0)/\theta_0'(0)
\simeq-8.97\times10^{-67}$.
The absolute magnitude of this response discrepancy
is $1.72\times10^{-14}$ of the local velocity
response. Its numerical stability is checked by
varying the momentum quadrature resolution
and integration range.

This result quantifies the mismatch between
the two projected source prescriptions.
It is not an independent calculation of the
negative-energy and particle--antiparticle
coherences omitted by the projection.
In particular, it is distinct from the
retarded-current comparison bound above.
The two estimates test different aspects of
the reduced calculation and must not be
identified with one another.

Finally, we test the numerical
stability of the calculations.
For the local response, the
Gauss--Legendre quadrature is varied
from 100 to 380 nodes, with the
upper momentum boundary varied from
$x=32$ to $64$.
The relative deviations of the
present-day susceptibility and local
energy-transfer rate from the
260-node, $x\leq48$ reference
are below approximately $10^{-8}$.

For the nonlocal-current comparison,
we increase the time grid from
401 to 1201 points and the momentum
quadrature from 12 to 36 nodes per
subinterval.
We use 11 momentum subintervals
over $0\leq x\leq80$.
Across these refinements, the
present-day fractional-velocity
difference envelope changes by
approximately $1.70\times10^{-4}$
in relative terms.
Further time-resolution and
momentum-cutoff convergence tests
are presented in
Appendix~\ref{app:nonlocal_comparison}.

The combined checks support the
accuracy of the local reactive
velocity response for the specified
benchmark within the linearized,
collisionless positive-energy
two-state calculation.
The finite-time comparison and
the Ward-projection test provide
complementary consistency checks,
rather than interchangeable
estimates of the same correction.

We also examine the sensitivity of the backreaction
to the preparation epoch. Keeping the reference
masses, coupling scales, and $T_{\rm PD0}$ fixed,
we repeat the calculation for $a_i=0.001$,
$0.003$, $0.01$, and $0.03$.
At each epoch, we reconstruct the initial thermal
occupations and impose $\theta(a_i)=1$,
$\theta'(a_i)=0$, and $\rho_{12,qh}(a_i)=0$.
These choices define distinct preparation histories,
rather than different starting points along
a single scalar trajectory.

At $a=1$, the local fractional velocity response
ranges from $5.08\times10^{-54}$ to
$1.29\times10^{-52}$.
The largest numerically evaluated bound on
$|R_{\rm ret}-R_{\rm ad}|/|R_{\rm ad}|$
among these runs is $2.04\times10^{-14}$.
Thus, none of the four tested preparation histories
shows significant Majoron freezing within the
linearized, collisionless positive-energy two-state
calculation.

\section{Discussion and Conclusions}
\label{sec:discussion_conclusions}

The central question of this work is whether the influence of a hidden fermion sector on an ultralight Majoron can be computed from a microscopic current and a specified quantum state, without postulating a damping coefficient or an energy-transfer law.
We have addressed this question for a thermally prepared hidden pseudo-Dirac ensemble evolving in the collisionless linear-response regime. The connection between the operator Ward identity, the finite-time Majorana coherence, and the resulting cosmological
backreaction is the main result of our analysis. For the benchmark studied here, the response is predominantly reactive and does not produce significant freezing.

The hidden Dirac fermion splits into two nearly degenerate Majorana mass eigenstates when a small Dirac-number-violating mass is introduced. The pseudo-Dirac vector current then becomes off diagonal
in the Majorana mass basis, as shown in
Eq.~\eqref{eq:III_J_majorana}. Its divergence is fixed by the exact
operator Ward identity in Eq.~\eqref{eq:III_ward_FRW}, with the mass splitting $\delta m=m_1-m_2>0$. With positive $Q_{\rm micro}$ defined as energy transferred from the Majoron to the hidden sector,
Eq.~\eqref{eq:Q_ward_framework} identifies the microscopic operator responsible for energy exchange. It does not determine the expectation value of that operator, which depends on the evolving
fermionic state, nor does it imply irreversible dissipation.
This distinction is essential when assessing a proposed freezing mechanism.

The spinor calculation provides an independent check of the current normalization. The projected vector and scalar matrix elements obey
the fixed-time Ward relation in Eq.~\eqref{eq:III_AB_Ward}.
The comoving momentum $q$ is held fixed during free streaming, while the physical momentum is $p=q/a$. The exact operator identity also remains valid in an evolving spinor basis. In the complete spinor space, the geometric connection accounts for the basis time
dependence. The residual term proportional to $\dot A_q$ in the positive-energy projection must not be interpreted as new symmetry breaking or irreversible dissipation.

We specify the fermionic state at a finite preparation time $t_i$.
It has Fermi--Dirac occupations, vanishing chemical potentials, and no off-diagonal coherence in the Majorana mass basis.
The Majoron initially satisfies $\dot\phi(t_i)=0$, so the
initial derivative drive vanishes and the density matrix commutes with the free mass Hamiltonian. After preparation, the unperturbed occupations remain fixed at each comoving momentum. The moving Majoron generates coherence dynamically through the collisionless
retarded evolution in Sec.~\ref{sec:coherent_kinetics}.
These conditions do not impose thermal equilibrium at later times.

The finite preparation time has a physical consequence. The exact retarded coherence retains an initial boundary response even when $|\Delta E_q|\gg H$. The boundary term proportional to the initial
drive vanishes because $\dot\phi(t_i)=0$, but the next term generally does not: the initial acceleration $\ddot\phi(t_i)$ can be nonzero.
Thus the local adiabatic particular solution need not satisfy the exact initial condition by itself. A fast coherence frequency does
not justify dropping the boundary response or the remaining memory. Their contributions must be checked rather than removed by an assumed damping or phase-averaging prescription.

The leading local current response is given in
Eq.~\eqref{eq:IV_Jlocal}. It depends on the free-streaming
susceptibility $\chi_{\rm PD}^{\rm free}$, evaluated using the initial occupation difference rather than an instantaneous thermal
distribution. The corresponding Majoron equation,
Eq.~\eqref{eq:IV_phieom_local}, contains a small reactive inertia
correction $\epsilon=\chi_{\rm PD}^{\rm free}/f_{\rm eff}^2$.
Neither this correction nor the term proportional to $\dot\phi$
by itself establishes microscopic dissipative friction.
The collisionless system evolves unitarily, and the instantaneous energy exchange can change sign.

We evaluated this response on a prescribed FRW background for a subdominant Majoron with $m_\phi=10^{-33}\,\mathrm{eV}$ and
$f_\phi=10^9\,\mathrm{GeV}$. The derivative-coupling scale is $f_{\rm eff}=2.03\times10^{11}\,\mathrm{GeV}$, and the hidden pseudo-Dirac parameters are $m_D=10^{-5}\,\mathrm{eV}$ and
$\delta m=2\times10^{-10}\,\mathrm{eV}$. The thermal state is prepared at $z_i=99$ and then free streams. For relativistic modes, the resulting susceptibility scales approximately as
$\chi_{\rm PD}^{\rm free}\propto a^{-4}$. This scaling follows from the preserved comoving occupations and differs from a continually rethermalized ensemble.

At the present epoch, the local calculation gives
$\epsilon_0=3.65\times10^{-54}$ and
$Q_{\rm micro}^{\rm ad}(t_0)=-3.85\times10^{-119}\,\mathrm{eV}^5$.
The dimensionless transfer rate is
$Q_{\rm micro}^{\rm ad}(t_0)/[3H_0\rho_\phi(t_0)]
=-2.10\times10^{-56}$, while the fractional velocity
change is $\delta\dot\phi_{\rm ad}(t_0)/\dot\phi_0(t_0)
=+5.23\times10^{-53}$.
Here $\dot\phi_0$ is the velocity of the uncoupled Majoron
solution, and $\delta\dot\phi_{\rm ad}$ denotes the local
adiabatic velocity perturbation. Because $\dot\phi_0(t_0)<0$,
the positive final ratio represents a minute increase in the
magnitude of the rolling velocity, not freezing. The local
energy-transfer rate also reverses sign during the evolution.
These results demonstrate negligible local reactive backreaction for this benchmark; they do not exclude different behavior at other parameter values.

We also tested whether the finite-time memory changes this
cosmological conclusion. The modewise inequality derived from the retarded coherence includes the initial boundary transient and the remaining memory integral. Its momentum-integrated current envelope is propagated through a nonnegative comparison
system for the scalar perturbation. As summarized in
Eq.~\eqref{eq:V_retarded_velocity_bound}, the present-day local response is $R_{\rm ad}=+5.23\times10^{-53}$. The
nonlocal--local difference satisfies
$|R_{\rm ret}-R_{\rm ad}|\lesssim3.54\times10^{-69}$,
corresponding to a relative bound of
$|R_{\rm ret}-R_{\rm ad}|/|R_{\rm ad}|
\lesssim6.78\times10^{-17}$.
Here $R=\delta\dot\phi/\dot\phi_0$ denotes the fractional
velocity response, with the subscripts ``ret'' and ``ad''
referring to the finite-time retarded and local adiabatic
responses, respectively. 

The estimate bounds the difference
in their present-day velocity responses within the linearized collisionless positive-energy two-state calculation.
It is obtained by propagating a current-difference inequality,
not by directly resolving all rapid coherence oscillations
in two independent coupled FRW solutions. The derivation and
numerical convergence tests are given in
Appendix~\ref{app:nonlocal_comparison}. The numerical
comparison is convergence-tested rather than interval-certified.
Within the linearized collisionless positive-energy
two-state calculation, the evaluated bound shows
that the finite-time boundary and memory contributions do not produce significant freezing for this benchmark.

Several limitations remain distinct from this verified
retarded-current comparison. First, the comparison above concerns the
current-induced scalar velocity response, not a pointwise
bound on the full instantaneous energy-transfer rate
$Q_{\rm micro}(t)$. Second, the evolving positive-energy
projection has a Ward-source correction distinct from the
retarded-versus-local current difference. Its contribution
to the present fractional velocity response is approximately
$-8.97\times10^{-67}$, or $1.72\times10^{-14}$ of the
local response in absolute relative magnitude, as shown in
Sec.~\ref{subsec:V_checks}. Neither estimate includes a
complete evolution of negative-energy or particle--antiparticle coherences. The exact operator Ward identity is unaffected by these projection limitations.

The preparation history is another limitation. Our initial
thermal ensemble could be established by earlier hidden-sector interactions, but those processes are not calculated here. Maintained equilibrium, scattering, or dephasing after $t_i$ would require explicit collision operators and would define a different dynamical system. We have not assumed such terms.

The scalar potential parameters and $f_{\rm eff}$ are treated as independent low-energy inputs. The Planck-suppressed scalar
operator and the illustrative derivative portal provide an
effective-theory construction, not a complete ultraviolet
completion or a full assessment of hidden-sector cosmological constraints. We also prescribe the FRW expansion rather than solve the fully coupled Einstein--Majoron--fermion system. 
The Majoron need not account for the observed dark-energy
density in this calculation.

The relevant limits must also be kept separate. The
positive-energy reduction requires sufficiently small mixing
with negative-energy states, including $H\ll E_{\alpha,q}$.
The local adiabatic expansion additionally requires the source and level splitting to vary slowly compared with the coherence frequency in the momentum region dominating the response.
The finite-time retarded solution approaches the conserved
Dirac limit smoothly as $\delta m\to0$ at fixed observation
time. By contrast, the formal ratio $D_q/\Delta E_q$ in the
static susceptibility can remain finite in that limit.
Consequently, the zero-splitting limit and the static
low-frequency limit cannot generally be interchanged.
The local formula must not be extrapolated into
$|\Delta E_q|(t-t_i)\lesssim1$ without checking the
finite-time response.

In summary, the microscopic Ward identity specifies the
Majoron--fermion energy-transfer operator, while the prepared quantum state and its causal retarded evolution determine the corresponding response. The finite-time analysis shows why a local reactive approximation requires control of boundary and
memory contributions. The independently evaluated comparison
bound confirms that, for the benchmark studied here, these
contributions do not produce an appreciable change in the
Majoron velocity. This is a benchmark-specific result, not a
general exclusion of freezing in interacting scalar--fermion
systems. A self-consistent treatment of collisions, persistent thermalization, and the full spinor density matrix, together with a complete ultraviolet construction, remains a distinct subject for future work.

\appendix

\section{Scalar potential and projection onto the physical Majoron direction}
\label{app:scalar_potential_majoron}

In this appendix we present the scalar potential and show explicitly how the
physical Majoron direction and its periodic potential arise for the charge
assignment specified in Sec.~\ref{subsec:majoron_sector}.
The most general renormalizable scalar potential involving the Standard Model
Higgs doublet \(H\) and the two singlet scalars \(\Phi_1,\Phi_2\), consistent
with the gauged \(U(1)_{B-L}\) symmetry, is
\begin{align}
V_{\rm ren}
=&
-\mu_H^2 H^\dagger H
-\mu_1^2 \Phi_1^\dagger\Phi_1
-\mu_2^2 \Phi_2^\dagger\Phi_2
\nonumber\\
&+
\lambda_H(H^\dagger H)^2
+\lambda_1(\Phi_1^\dagger\Phi_1)^2
+\lambda_2(\Phi_2^\dagger\Phi_2)^2
\nonumber\\
&+
\lambda_{12}(\Phi_1^\dagger\Phi_1)(\Phi_2^\dagger\Phi_2)
+\lambda_{H1}(H^\dagger H)(\Phi_1^\dagger\Phi_1)
\nonumber\\
&+\lambda_{H2}(H^\dagger H)(\Phi_2^\dagger\Phi_2).
\label{eq:app_Vren}
\end{align}
Here $\mu_H^2$, $\mu_1^2$, and $\mu_2^2$
are the quadratic scalar mass parameters.
The coefficients $\lambda_H$, $\lambda_1$,
and $\lambda_2$ are the scalar self-couplings,
while $\lambda_{12}$, $\lambda_{H1}$,
and $\lambda_{H2}$ are the quartic portal
couplings between different scalar fields.
For the benchmark charges in Sec.~\ref{subsec:majoron_sector}, no renormalizable
phase-sensitive operator involving both \(\Phi_1\) and \(\Phi_2\) is allowed
by \(U(1)_{B-L}\). Thus \(V_{\rm ren}\) fixes the radial vacuum expectation values but does not generate a potential for the physical Majoron direction.

After electroweak and \(U(1)_{B-L}\) symmetry breaking,
\begin{equation}
H=
\frac{1}{\sqrt2}
\begin{pmatrix}
0\\ v+h
\end{pmatrix},
\qquad
\Phi_j=
\frac{1}{\sqrt2}(v_j+\rho_j)e^{i\theta_j},
\label{eq:app_ssb_fields}
\end{equation}
the stationary conditions are
\begin{align}
\mu_H^2
&=
\lambda_H v^2
+\frac12\lambda_{H1}v_1^2
+\frac12\lambda_{H2}v_2^2,
\nonumber\\
\mu_1^2
&=
\lambda_1 v_1^2
+\frac12\lambda_{12}v_2^2
+\frac12\lambda_{H1}v^2,
\nonumber\\
\mu_2^2
&=
\lambda_2 v_2^2
+\frac12\lambda_{12}v_1^2
+\frac12\lambda_{H2}v^2.
\label{eq:app_tadpole_conditions}
\end{align}
In the CP-even scalar basis \((h,\rho_1,\rho_2)\), the tree-level mass matrix
is
\begin{equation}
\mathcal M^2_{\rm even}
=
\begin{pmatrix}
2\lambda_H v^2 & \lambda_{H1}vv_1 & \lambda_{H2}vv_2\\
\lambda_{H1}vv_1 & 2\lambda_1v_1^2 & \lambda_{12}v_1v_2\\
\lambda_{H2}vv_2 & \lambda_{12}v_1v_2 & 2\lambda_2v_2^2
\end{pmatrix}.
\label{eq:app_even_mass_matrix}
\end{equation}
The CP-even radial modes are assumed to be heavy compared with the late-time Majoron dynamics and are integrated out in the low-energy effective theory.

A small potential for the physical Majoron can be generated by a Planck-suppressed operator that is invariant under the gauged
\(U(1)_{B-L}\) symmetry but explicitly breaks the accidental global phase
symmetry of the renormalizable singlet sector.  A gauge-invariant operator
aligned with the physical Majoron direction is
\begin{equation}
V_{\rm br}
=
-
\left[
\kappa
\frac{\Phi_1^{m}(\Phi_2^\dagger)^{n}}
{M_{\rm Pl}^{m+n-4}}
+\mathrm{h.c.}
\right],
\label{eq:app_Vbr_operator}
\end{equation}
where $\kappa$ is a dimensionless complex coefficient,
$m$ and $n$ are positive integers specifying the powers
of the scalar fields, and
$M_{\rm Pl}=1.22\times10^{19}\,\mathrm{GeV}$ is the
non-reduced Planck mass. Here h.c. denotes the
Hermitian conjugate.
This operator is gauge invariant because
\begin{equation}
m Q_1- n Q_2=0.
\label{eq:app_operator_gauge_invariance}
\end{equation}
Substituting Eq.~\eqref{eq:app_ssb_fields} into
Eq.~\eqref{eq:app_Vbr_operator} and integrating out the radial modes gives
\begin{align}
V_{\rm br}(\phi)
&=-\frac{\kappa}{M_{\rm Pl}^{m+n-4}}
\left(\frac{v_1}{\sqrt2}\right)^m
\left(\frac{v_2}{\sqrt2}\right)^n
\nonumber\\
&\quad\times e^{i(m\theta_1-n\theta_2)}
+\mathrm{h.c.}
\nonumber\\
&=-\Lambda_\phi^4
\cos\left(\frac{\phi}{f_\phi}+\delta\right),
\label{eq:app_Vbr_cosine}
\end{align}
where
\begin{equation}
\delta=\arg\kappa,
\qquad
\Lambda_{\phi}^4
\equiv
\frac{2|\kappa|}{2^{(m+n)/2}}
\frac{v_1^{m}v_2^{n}}
{M_{\rm Pl}^{m+n-4}}.
\label{eq:app_LambdaJ_general}
\end{equation}
Thus the renormalizable scalar potential generates the
\(U(1)_{B-L}\)-breaking vacuum and leaves an uneaten physical phase direction,
while the Planck-suppressed operator in
Eq.~\eqref{eq:app_Vbr_operator} gives the physical Majoron a small periodic potential.

\section{Origin of the pseudo-Dirac mass structure}
\label{app:hidden_pseudo_dirac_origin}

In the main text we use the minimal realization in which the hidden
pseudo-Dirac fermions are singlets under the gauged \(U(1)_{B-L}\),
\begin{equation}
Q_{B-L}(N)=Q_{B-L}(S)=0 .
\end{equation}
Here $Q_{B-L}(X)$ denotes the charge of a field
$X$ under the gauged $U(1)_{B-L}$ symmetry.
This choice separates the hidden pseudo-Dirac fermion from the ordinary right-handed neutrinos responsible for the observed active-neutrino
masses. It also avoids direct \(B-L\) gauge interactions of the hidden reservoir. The Majoron coupling to the hidden sector is therefore not a
minimal gauge interaction induced by a \(B-L\) charge assignment. It is instead treated as an effective hidden-sector portal to the pseudo-Dirac number current.

Because \(N\) and \(S\) are \(B-L\) singlets, the pseudo-Dirac mass terms used in the main text are gauge invariant already at the effective-field-theory level. The most general symmetric two-state mass
structure relevant for the hidden reservoir is
\begin{equation}
{\cal L}_{\rm mass}^{\rm PD}
=
-m_D NS
-\frac12 \mu_N NN
-\frac12 \mu_S SS
+{\rm h.c.}
\end{equation}
Here $\mu_N$ and $\mu_S$ are the Majorana
mass parameters associated with the
left-handed Weyl fields $N$ and $S$,
respectively. The pseudo-Dirac limit corresponds to
\begin{equation}
|\mu_N|,|\mu_S|\ll m_D .
\end{equation}
In the limit \(\mu_N,\mu_S\to0\), the hidden sector recovers a conserved
hidden pseudo-Dirac number symmetry \(U(1)_{\rm PD}\), under which \(N\)
and \(S\) carry opposite hidden charges. The small Majorana entries softly break
this hidden number symmetry.
They are technically natural in the
usual 't Hooft sense, provided the
remaining interactions respect
$U(1)_{\rm PD}$ in the limit
$\mu_N,\mu_S\to0$.
In this limit, setting the Majorana
entries to zero enhances the symmetry
of the hidden sector.

For the benchmark model used in the main text, we take the
exchange-symmetric limit
\begin{equation}
\mu_N=\mu_S\equiv \mu .
\end{equation}
This can be viewed as the minimal limit selected by an approximate
hidden exchange symmetry \(N\leftrightarrow S\). In this limit the mass Lagrangian becomes
\begin{equation}
{\cal L}_{\rm mass}^{\rm PD}
=
-m_D NS
-\frac12\mu NN
-\frac12\mu SS
+{\rm h.c.}
\end{equation}
The corresponding mass matrix in the \((N,S)\) basis is
\begin{equation}
{\cal M}=
\begin{pmatrix}
\mu & m_D \\
m_D & \mu
\end{pmatrix}.
\end{equation}
It is diagonalized by the symmetric and antisymmetric combinations
\begin{equation}
N_{1}=\frac{N+S}{\sqrt2},
\qquad
N_{2}=\frac{i(N-S)}{\sqrt2}.
\end{equation}
The factor of \(i\) is a phase convention chosen so that both physical Majorana masses are positive. With this convention,
\(m_{1}=m_D+\mu\) and \(m_{2}=m_D-\mu\). This is the mass structure used in the main text. The corresponding mass splitting is defined in
Eq.~\eqref{eq:hidden_mass_splitting}.
The pseudo-Dirac number current is introduced in
Eq.~\eqref{eq:III_J_majorana}, and its two-state
projection is derived in Sec.~\ref{subsec:ward_spinor_projection}.

\section{Four-component Ward consistency}
\label{app:four_component_ward}

We verify the Ward relation in the complete
four-component mode space. At fixed comoving
momentum $q=ap$ and helicity $h=\pm1$, the free
Hamiltonian can be written as
\begin{align}
\mathsf H_{0,qh}
={}&-hp\,I_2\otimes\sigma_z
+\bar m\,I_2\otimes\sigma_x
\nonumber\\
&+\frac{\delta m}{2}
\sigma_z\otimes\sigma_x,
\label{eq:appC_free_H}
\end{align}
where $\bar m=(m_1+m_2)/2$ is the mean
Majorana mass.

For a homogeneous Majoron background,
$\mathsf H_{qh}=\mathsf H_{0,qh}
-\dot\phi\,\mathsf Q/f_{\rm eff}$,
with $\mathsf Q=\sigma_y\otimes I_2$ and
$\mathsf S=\sigma_x\otimes\sigma_x$.
Since $[\mathsf Q,\mathsf Q]=0$, we obtain
\begin{align}
i[\mathsf H_{qh},\mathsf Q]
&=i[\mathsf H_{0,qh},\mathsf Q]
\nonumber\\
&=\frac{i\delta m}{2}
[\sigma_z,\sigma_y]\otimes\sigma_x
\nonumber\\
&=\delta m\,\mathsf S.
\label{eq:appC_commutator}
\end{align}

Let $\mathcal R_{qh}$ denote the complete
equal-time fermionic covariance matrix,
including both energy sectors.
Its collisionless evolution satisfies
$i\dot{\mathcal R}_{qh}=[\mathsf H_{qh},\mathcal R_{qh}]$.
Consequently,
\begin{align}
\frac{d}{dt}\operatorname{Tr}
(\mathcal R_{qh}\mathsf Q)
&=i\operatorname{Tr}
\bigl(\mathcal R_{qh}
[\mathsf H_{qh},\mathsf Q]\bigr)
\nonumber\\
&=\delta m\,
\operatorname{Tr}(\mathcal R_{qh}\mathsf S).
\label{eq:appC_trace_Ward}
\end{align}

To connect the full mode-space matrix relation
with the field operators, let $\widehat\Xi_{qh}$
denote the four-component Nambu operator formed
from the two Majorana Weyl modes and their
charge-conjugate partners. More explicitly, we write
$\widehat\Xi_{qh}=(\widehat\chi_{1,qh},
\widehat\chi_{2,qh})^T$, where
$\widehat\chi_{\alpha,qh}$ contains a Weyl mode
and its charge-conjugate partner in the reduced
helicity space. The Nambu components obey the
canonical fermionic anticommutation relations,
with the charge-conjugation redundancy understood.
The field rescaling
$\psi_\alpha=a^{3/2}\Psi_\alpha$,
used in Sec.~\ref{subsec:ward_spinor_projection},
removes the spin-connection dilution term.
With the same Majorana phase convention as in
Eq.~\eqref{eq:ND_definition}, the charge density
and the off-diagonal scalar bilinear are represented
by the Nambu quadratic forms
\begin{equation}
\begin{aligned}
\widehat{\mathcal J}_{qh}
&=\frac12:\widehat\Xi_{qh}^{\dagger}
\mathsf Q\widehat\Xi_{qh}:,\\
\widehat{\mathcal S}_{qh}
&=\frac12:\widehat\Xi_{qh}^{\dagger}
\mathsf S\widehat\Xi_{qh}:.
\end{aligned}
\label{eq:appC_nambu_bilinears}
\end{equation}
The factor $1/2$ removes the double counting of
particle and charge-conjugate modes in the Nambu
representation. The first operator reproduces
$J_{\rm PD}^0=:N^\dagger N-S^\dagger S:$
from Eq.~\eqref{eq:III_J_weyl}; the second is the
corresponding representation of
$:\bar\Psi_1\Psi_2:$.
Thus the matrices $\mathsf Q$ and $\mathsf S$
represent the same charge and scalar source that
occur in the field-level Ward identity, rather
than independently normalized one-particle observables.

With the index convention
$(\mathcal R_{qh})_{ab}
=\langle\widehat\Xi_{qh,b}^\dagger
\widehat\Xi_{qh,a}\rangle$, the two physical
densities are
\begin{equation}
\begin{aligned}
\langle J_{\rm PD}^0\rangle
&=\frac{1}{2a^3}\sum_h\!\int\!\frac{d^3q}{(2\pi)^3}
\operatorname{Tr}\!\left[\bigl(\mathcal R_{qh}
-\mathcal R^{\rm ref}_{qh}\bigr)\mathsf Q\right],\\
\langle:\bar\Psi_1\Psi_2:\rangle
&=\frac{1}{2a^3}\sum_h\!\int\!\frac{d^3q}{(2\pi)^3}
\operatorname{Tr}\!\left[\bigl(\mathcal R_{qh}
-\mathcal R^{\rm ref}_{qh}\bigr)\mathsf S\right].
\end{aligned}
\label{eq:appC_field_trace}
\end{equation}
Here $\mathcal R^{\rm ref}_{qh}$ is the
source-free, species-diagonal reference-vacuum
covariance implementing the normal ordering.
Because $\mathsf Q$ and $\mathsf S$ are both
off diagonal in Majorana-species space,
$\operatorname{Tr}(\mathcal R^{\rm ref}_{qh}\mathsf Q)
=\operatorname{Tr}(\mathcal R^{\rm ref}_{qh}\mathsf S)=0$
at each time. No time-dependent subtraction term
therefore enters the following identity.
Applying Eq.~\eqref{eq:appC_trace_Ward} to
Eq.~\eqref{eq:appC_field_trace}, while keeping
$q$ fixed in the time derivative, gives
\begin{equation}
\begin{aligned}
&\left(\partial_t+3H\right)
\langle J_{\rm PD}^0\rangle\\
&\quad=\frac{\delta m}{2a^3}\sum_h
\int\!\frac{d^3q}{(2\pi)^3}
\operatorname{Tr}(\mathcal R_{qh}\mathsf S)\\
&\quad=\delta m
\langle:\bar\Psi_1\Psi_2:\rangle.
\end{aligned}
\label{eq:appC_FRW_trace_Ward}
\end{equation}
This is precisely the homogeneous field-operator
Ward relation in Eq.~\eqref{eq:III_ward_FRW}.
The factor $3H$ comes only from differentiating
the physical volume factor $a^{-3}$.
No positive-energy truncation or expansion in
$\delta m$ was made in this correspondence.

As a check of the normalization, consider the
charge-conjugation-symmetric state without
positive/negative-energy pair coherence.
In the complete instantaneous energy basis,
its covariance has blocks $\rho_{qh}$ and
$I_2-\rho_{qh}^{T}$, while the diagonal
blocks of either bilinear matrix obey
$\mathsf Q_{e,--}=-\mathsf Q_{e,++}^{T}$ and
$\mathsf S_{e,--}=-\mathsf S_{e,++}^{T}$.
Since both matrices are traceless,
\begin{equation}
\begin{aligned}
\frac12\operatorname{Tr}
(\mathcal R_{e,qh}\mathsf Q_e)
&=\operatorname{tr}(\rho_{qh}A_q\sigma_y)
=-2A_q\operatorname{Im}\rho_{12,qh},\\
\frac12\operatorname{Tr}
(\mathcal R_{e,qh}\mathsf S_e)
&=\operatorname{tr}(\rho_{qh}B_q\sigma_x)
=2B_q\operatorname{Re}\rho_{12,qh}.
\end{aligned}
\label{eq:appC_positive_check}
\end{equation}
Restoring the common $a^{-3}$ momentum integral
recovers Eqs.~\eqref{eq:III_J_density_matrix}
and \eqref{eq:III_scalar_density_matrix}.
This last step is a check of the projected limit;
the full Ward relation above does not assume
that the pair-coherence blocks vanish.
The explicit cancellation of time-dependent
basis contributions is verified below.

To verify the cancellation explicitly, choose the real
instantaneous eigenspinors
$u_{\alpha}=(\cos\theta_{\alpha},\sin\theta_{\alpha})^T$ and
$v_{\alpha}=(-\sin\theta_{\alpha},\cos\theta_{\alpha})^T$,
where $\theta_{\alpha}$ parametrizes the instantaneous
spinor eigenbasis and $v_{\alpha}$ is the normalized
negative-energy eigenspinor.
At fixed comoving momentum $q$, their time dependence
is determined by
\begin{equation}
\dot\theta_{\alpha}
=-\frac{hH m_{\alpha}p}{2E_{\alpha,q}^2},
\qquad p=\frac{q}{a}.
\label{eq:appC_theta_dot}
\end{equation}
Defining $s_q=\sin(\theta_1-\theta_2)$, we have
$A_q=\cos(\theta_1-\theta_2)$ and
$\dot A_q=-s_q(\dot\theta_1-\dot\theta_2)$,
where $s_q$ measures the relative rotation
of the two instantaneous spinor bases.

In the complete basis $(1+,2+,1-,2-)$, the
positive-negative-energy blocks of the current and
geometric connection are
\begin{equation}
\begin{aligned}
(\mathsf Q_e)_{+-}&=-is_q\sigma_x,\\
(\mathsf Q_e)_{-+}&=is_q\sigma_x,\\
(\Omega_{qh})_{+-}&=-D_\theta,\\
(\Omega_{qh})_{-+}&=D_\theta,
\end{aligned}
\label{eq:appC_connection_blocks}
\end{equation}
where $D_\theta=\operatorname{diag}
(\dot\theta_1,\dot\theta_2)$.
Consequently,
\begin{equation}
\begin{aligned}
{}[\mathsf Q_e,\Omega_{qh}]_{++}
&=is_q[D_\theta,\sigma_x]\\
&=\dot A_q\sigma_y.
\end{aligned}
\label{eq:appC_connection_commutator}
\end{equation}
The positive-energy block therefore satisfies
\begin{equation}
\begin{aligned}
&(\dot{\mathsf Q}_e)_{++}
+i[-i\Omega_{qh},\mathsf Q_e]_{++}\\
&\qquad=\dot A_q\sigma_y
-\dot A_q\sigma_y=0.
\end{aligned}
\label{eq:appC_connection_cancellation}
\end{equation}
Thus the apparent projection residual is cancelled
exactly by the positive-negative-energy connection
blocks in the complete spinor description.

\section{Illustrative gauge-invariant origin of the derivative portal}
\label{app:derivative_portal_origin}

The main calculation uses the matter action
\eqref{eq:effective_action} without assuming a specific ultraviolet
completion. Here we collect a possible origin of its derivative portal.
An illustrative gauge-invariant origin of the derivative
coupling is the dimension-six operator
\begin{equation}
\mathcal L_{\rm portal}^{(6)}
=
\frac{J^\mu_{\rm PD}}{\Lambda^2}
\sum_{j=1}^{2}
c_j\,i\Phi_j^\dagger
\overleftrightarrow{D_\mu}\Phi_j,
\label{eq:appD_portal}
\end{equation}
where $\Lambda$ is the effective suppression scale,
$c_j$ are dimensionless portal coefficients,
and $D_\mu=\partial_\mu-i g_{BL}Q_jX_\mu$
is the gauge-covariant derivative.
After symmetry breaking, this operator generates
a derivative coupling of the physical Majoron
for suitable coefficients.

Let us make this origin explicit.
Using the scalar parameterization in
Eq.~\eqref{eq:Phi_phases}, we introduce
$a_j\equiv v_j\theta_j$ as the dimensionful
phase coordinates. After removing the heavy
radial fluctuations, the gauge-invariant
currents satisfy
\begin{equation}
i\Phi_j^\dagger
\overleftrightarrow{D_\mu}\Phi_j
=
-v_j\partial_\mu a_j
+g_{BL}Q_jv_j^2X_\mu.
\label{eq:appD_scalar_current}
\end{equation}
The gauge-invariant phase orthogonal to the
eaten Goldstone mode is
\begin{equation}
\phi=
\frac{Q_2v_2a_1-Q_1v_1a_2}{F},
\qquad
F=\sqrt{Q_1^2v_1^2+Q_2^2v_2^2}.
\label{eq:appD_majoron_direction}
\end{equation}
In the minimal two-phase sector, with no additional
light-state mixing, projection onto the physical
Majoron yields
\begin{equation}
\frac{1}{f_{\rm eff}}
=
\frac{v_1v_2}{F\Lambda^2}
\left|c_2Q_1-c_1Q_2\right|,
\label{eq:appD_feff}
\end{equation}
where an overall interaction sign can be absorbed
into the definition of the Majoron field.
For the illustrative charge assignment
$Q_1=2$, $Q_2=15$, and $c_1=0$, this reduces to
\begin{equation}
f_{\rm eff}
=
\frac{\Lambda^2}{C_2v_2},
\qquad
C_2=\frac{2|c_2|v_1}{F}.
\label{eq:appD_feff_example}
\end{equation}
Thus the derivative interaction follows directly
from the symmetry-breaking projection of the
gauge-invariant portal.
For \(|c_2|=1\), the benchmark corresponds to \(\Lambda\simeq2.02\times10^{10}\) GeV, below \(v_1\). Thus, this construction is illustrative and does not establish a controlled ultraviolet completion across both symmetry-breaking scales.

\section{Finite-time nonlocal comparison and numerical error control}
\label{app:nonlocal_comparison}

This appendix supplies the derivation and numerical
implementation of the current-difference envelope used
in Sec.~\ref{subsec:V_checks}. We work with the
collisionless, linearized positive-energy two-state
system and the same prepared fermion distribution
as in Sec.~\ref{subsec:IV_initial_value}.
The uncoupled Majoron solution $\phi_0(t)$ and the
prescribed FRW expansion are used consistently
in both currents. The comparison controls the
finite-time memory and initial boundary response
within this system; no oscillatory phase average
is assumed.

\subsection{Retarded identity and modewise envelope}

With the source $F_{qh}$ and frequency $\omega_q$
introduced in Eq.~\eqref{eq:IV_omega_s}, the
finite-time coherence obeys
\begin{equation}
\rho_{12,qh}^{\rm ret}(t)
=\int_{t_i}^{t}dt'\,F_{qh}(t')
 e^{-i\Theta_q(t,t')},
\label{eq:appE_retarded}
\end{equation}
where $\Theta_q(t,t')=\int_{t'}^t\omega_q(s)ds$
is the accumulated coherence phase.
For the prepared condition $\dot\phi_0(t_i)=0$,
we have $F_{qh}(t_i)=0$.
Using
$\partial_{t'}e^{-i\Theta_q(t,t')}
=i\omega_q(t')e^{-i\Theta_q(t,t')}$,
the first integration by parts gives
\begin{equation}
\begin{aligned}
\rho_{12,qh}^{\rm ret}(t)
={}&-\frac{iF_{qh}(t)}{\omega_q(t)}\\
&+i\int_{t_i}^{t}dt'\,
 e^{-i\Theta_q(t,t')}
 \frac{d}{dt'}\left(\frac{F_{qh}}{\omega_q}\right).
\end{aligned}
\label{eq:appE_first_parts}
\end{equation}
In the next integration by parts, use $L_{qh}$ from
Eq.~\eqref{eq:V_L_def} to recover the exact identity
in Eq.~\eqref{eq:V_retarded_parts}.
In particular,
\begin{equation}
L_{qh}(t_i)
=\frac{A_q(t_i)D_{qh,i}}
{f_{\rm eff}\omega_q^2(t_i)}\ddot\phi_0(t_i),
\label{eq:appE_boundary}
\end{equation}
where the initial acceleration follows from
$\ddot\phi_0(t_i)=-V_{,\phi}(\phi_i)$.
This quantity need not vanish even though
$F_{qh}(t_i)=0$.
Taking the imaginary part of
Eq.~\eqref{eq:V_retarded_parts}, and using
$|\operatorname{Im}z|\leq|z|$, gives the
modewise inequality in Eq.~\eqref{eq:V_modewise_bound}.
The bound uses absolute values before integrating
in time; it therefore does not rely on cancellation
between rapidly oscillating phases.

\subsection{From the coherence envelope to a current bound}

At fixed comoving momentum, define the nonnegative
modewise remainder
\begin{equation}
M_{qh}(t)=|L_{qh}(t_i)|
+\int_{t_i}^{t}dt'\,|\dot L_{qh}(t')|,
\label{eq:appE_M}
\end{equation}
where $M_{qh}$ bounds the imaginary-coherence
difference in Eq.~\eqref{eq:V_modewise_bound}.
The current expression in
Eq.~\eqref{eq:III_J_density_matrix} then gives
\begin{equation}
\begin{aligned}
|J_{\rm ret}^0-J_{\rm ad}^0|
\leq{}&\frac{2}{a^3}\sum_h
\int\frac{d^3q}{(2\pi)^3}\,A_q(t)M_{qh}(t).
\end{aligned}
\label{eq:appE_current_bound}
\end{equation}
For an isotropic, helicity-independent prepared
state, let $x=q/T_{\rm PD0}$, as in
Sec.~\ref{subsec:V_susceptibility}.
The corresponding dimensionless envelope is
\begin{equation}
\begin{aligned}
\beta(\mathcal N)
={}&\frac{2T_{\rm PD0}^3}
{\pi^2a^3f_{\rm eff}f_\phi H}
\int_0^\infty dx\,x^2 A_q(\mathcal N)\\
&\quad\times\left[
|L_{qh}(\mathcal N_i)|
+\int_{\mathcal N_i}^{\mathcal N}
 d\mathcal N'\,|L_{qh}'(\mathcal N')|
\right],
\end{aligned}
\label{eq:appE_beta}
\end{equation}
where a prime denotes $d/d\mathcal N$ and
the helicity dependence is identical for both states.
In deriving Eq.~\eqref{eq:appE_beta}, we used
$|\dot L_{qh}|dt=|L_{qh}'|d\mathcal N$.
Thus $|\Delta j|\leq\beta$, with $j$ defined
in Eq.~\eqref{eq:V_j_beta}.

\subsection{Nonnegative comparison system for the Majoron response}

Let $\delta\theta_{\rm ret}$ and
$\delta\theta_{\rm ad}$ be the first-order
scalar responses driven, respectively, by
$j_{\rm ret}$ and $j_{\rm ad}$ along the
same uncoupled solution $\theta_0$.
Define
$\Delta\theta=\delta\theta_{\rm ret}
-\delta\theta_{\rm ad}$ and
$\Delta j=j_{\rm ret}-j_{\rm ad}$.
The scalar equation and the current definition
imply
\begin{equation}
\begin{aligned}
\Delta\theta''+(3+h_{\mathcal N})\Delta\theta'
+K(\mathcal N)\Delta\theta
=-\Delta j'-(3+h_{\mathcal N})\Delta j,
\end{aligned}
\label{eq:appE_delta_theta}
\end{equation}
where
$K(\mathcal N)=(m_\phi^2/H^2)\cos\theta_0$
is the linearized Majoron restoring coefficient.
The current-derivative source is removed by
introducing
\begin{equation}
\begin{aligned}
u&=\Delta\theta'+\Delta j,\\
\Delta\theta'&=u-\Delta j,\\
u'&=-(3+h_{\mathcal N})u-K\Delta\theta.
\end{aligned}
\label{eq:appE_u}
\end{equation}
Here $u$ is an auxiliary response variable;
it contains the difference between the scalar
velocity perturbations and the normalized
current responses.
At preparation the retarded and leading local
currents vanish because $\dot\phi_0(t_i)=0$.
We also impose equal initial scalar positions
and velocities, so $\Delta\theta(\mathcal N_i)
=u(\mathcal N_i)=0$.

Introduce nonnegative comparison variables
$X(\mathcal N)$ and $Y(\mathcal N)$ satisfying
\begin{equation}
\begin{aligned}
X'&=Y+\beta,\\
Y'&=-(3+h_{\mathcal N})Y+|K|X,\\
X(\mathcal N_i)&=Y(\mathcal N_i)=0.
\end{aligned}
\label{eq:appE_comparison}
\end{equation}
Here $X$ and $Y$ are upper envelopes for
$|\Delta\theta|$ and $|u|$, respectively.
Since $3+h_{\mathcal N}>0$ for the prescribed
expanding $\Lambda$CDM background,
variation of constants in Eq.~\eqref{eq:appE_u}
and $|\Delta j|\leq\beta$ give
\begin{equation}
|\Delta\theta(\mathcal N)|\leq X(\mathcal N),
\qquad |u(\mathcal N)|\leq Y(\mathcal N).
\label{eq:appE_comparison_proof}
\end{equation}
These inequalities also follow by iterating the
corresponding positive integral kernels with
zero initial data.
Because $\Delta\theta'=u-\Delta j$,
\begin{equation}
\begin{aligned}
|R_{\rm ret}-R_{\rm ad}|
&=\frac{|\Delta\theta'|}{|\theta_0'|}\\
&\leq\frac{Y+\beta}{|\theta_0'|},
\qquad \theta_0'\ne0.
\end{aligned}
\label{eq:appE_velocity_bound}
\end{equation}
The ratio is evaluated at $a=1$, for which
$\theta_0'(0)\ne0$.
This construction bounds the scalar velocity
response without numerically differentiating
an extremely small current difference.

\subsection{Analytical derivatives and numerical implementation}

For reproducibility, all derivatives entering
$L_{qh}'$ are evaluated analytically, rather
than by finite differencing a rapidly oscillatory
retarded solution. Introduce
\begin{equation}
\begin{aligned}
k(\mathcal N)&=\frac{m_\phi^2}{H^2},\\
v(\mathcal N)&=\theta_0',\\
d(\mathcal N)&=3+h_{\mathcal N},\\
\alpha_q(\mathcal N)&=\frac{d\ln A_q}{d\mathcal N},\\
\nu_q(\mathcal N)&=\frac{d\ln\omega_q}{d\mathcal N},\\
b_q(\mathcal N)&=h_{\mathcal N}+\alpha_q-\nu_q,
\end{aligned}
\label{eq:appE_aux_defs}
\end{equation}
where $k$ is the dimensionless Majoron mass
term, $v$ the uncoupled field derivative,
$d$ the Hubble dilution coefficient,
$\alpha_q$ and $\nu_q$ the logarithmic
evolution rates of the overlap and coherence
frequency, and $b_q$ their combination entering
the differentiated source.
Define the positive (for the prepared ordering)
mode factor
\begin{equation}
\mathcal K_{qh}(\mathcal N)
=\frac{D_{qh,i}f_\phi}{f_{\rm eff}}
A_q\left(\frac{H}{\omega_q}\right)^2,
\label{eq:appE_Kfactor}
\end{equation}
where $\mathcal K_{qh}$ collects the slowly
varying factors in the first derivative
correction. Then
\begin{equation}
\begin{aligned}
L_{qh}&=\mathcal K_{qh}(v'+b_qv),\\
L_{qh}'&=\mathcal K_{qh}
\bigl[(2h_{\mathcal N}+\alpha_q-2\nu_q)
(v'+b_qv)\\
&\hspace{16mm}+v''+b_qv'+b_q'v\bigr].
\end{aligned}
\label{eq:appE_L_derivatives}
\end{equation}
Here the needed scalar derivatives obey
\begin{equation}
\begin{aligned}
v'&=-d v-k\sin\theta_0,\\
v''&=-d v'-h_{\mathcal N}'v
+2h_{\mathcal N}k\sin\theta_0
-k\cos\theta_0\,v.
\end{aligned}
\label{eq:appE_v_derivatives}
\end{equation}
The frequency derivatives follow from the
exact fixed-$q$ relation in
Eq.~\eqref{eq:V_omega_derivative}:
\begin{equation}
\begin{aligned}
\nu_q&=\frac{p^2}{E_{1,q}E_{2,q}},\\
\nu_q'&=-2\nu_q
+\nu_q p^2\left(\frac1{E_{1,q}^2}
+\frac1{E_{2,q}^2}\right).
\end{aligned}
\label{eq:appE_nu_derivatives}
\end{equation}
To evaluate the overlap without subtracting
nearly equal spinor components, introduce
\begin{equation}
\begin{aligned}
\eta_q&=\frac12\left[
\arctan\left(\frac{p}{m_1}\right)
-\arctan\left(\frac{p}{m_2}\right)\right],\\
A_q&=\cos\eta_q,\\
\eta_q'&=-\frac p2
\left(\frac{m_1}{E_{1,q}^2}
-\frac{m_2}{E_{2,q}^2}\right),\\
\eta_q''&=\frac p2
\left(\frac{m_1}{E_{1,q}^2}
-\frac{m_2}{E_{2,q}^2}\right)\\
&\quad-p^3\left(\frac{m_1}{E_{1,q}^4}
-\frac{m_2}{E_{2,q}^4}\right),
\end{aligned}
\label{eq:appE_overlap_derivatives}
\end{equation}
where $\eta_q$ is the helicity-independent
relative spinor angle. In the convention
of Appendix~\ref{app:four_component_ward},
the instantaneous spinor angles satisfy
$\theta_1-\theta_2=h\eta_q$.
The derivatives are taken at fixed
comoving momentum $q$, with $p=q/a$.
Consequently,
\begin{equation}
\begin{aligned}
\alpha_q&=-\tan\eta_q\,\eta_q',\\
\alpha_q'&=-\sec^2\eta_q\,(\eta_q')^2
-\tan\eta_q\,\eta_q'',\\
b_q'&=h_{\mathcal N}'+\alpha_q'-\nu_q'.
\end{aligned}
\label{eq:appE_alpha_derivatives}
\end{equation}
All quantities in Eq.~\eqref{eq:appE_L_derivatives}
are therefore specified by the uncoupled FRW
solution and the prepared distribution.

The numerical calculation integrates the
uncoupled scalar equation
\eqref{eq:V_background_theta}
from $\mathcal N_i=\ln(0.01)$ to $0$.
For the initial occupation difference we use
the stable thermal form in
Eq.~\eqref{eq:V_Dinit}, with the exact
energy splitting in Eq.~\eqref{eq:V_stable_delta}.
The momentum integral in Eq.~\eqref{eq:appE_beta}
is evaluated with Gauss--Legendre quadrature
on 11 equal subintervals over $0\leq x\leq80$.
The time integral of $|L_{qh}'|$ uses
cumulative trapezoidal integration on an
equally spaced $\mathcal N$ grid. A
shape-preserving interpolation of $\beta$
is supplied to an adaptive ODE solver for
Eq.~\eqref{eq:appE_comparison}.

\begin{table}[b]
\caption{Convergence of the numerical comparison
envelope at $a=1$ for the benchmark.
The quoted values are numerical evaluations of
the analytical comparison bound, not
interval-certified bounds.}
\label{tab:appE_convergence}
\begin{ruledtabular}
\begin{tabular}{ccc}
Time nodes & Nodes per panel &
$[Y(0)+\beta(0)]/|\theta_0'(0)|$ \\
\hline
401 & 12 & $3.54451871\times10^{-69}$ \\
801 & 24 & $3.54401171\times10^{-69}$ \\
1201 & 36 & $3.54391774\times10^{-69}$ \\
2401 & 48 & $3.54386135\times10^{-69}$ \\
6401 & 48 & $3.54384520\times10^{-69}$

\end{tabular}
\end{ruledtabular}
\end{table}

The convergence study in
Table~\ref{tab:appE_convergence} includes the
three resolutions discussed in Sec.~\ref{subsec:V_checks}
and tests two further time resolutions.
Across the coarsest and finest resolutions,
the relative change is approximately
$1.90\times10^{-4}$.
At fixed temporal and quadrature resolutions
(1201 time nodes with 36 quadrature nodes per
panel), increasing $x_{\max}$ from 64 to 96
changes the result by less than $10^{-9}$
in relative terms, confirming numerical
stability with respect to the momentum cutoff.
Using 8, 11, or 16 momentum panels gives
agreement in the displayed significant digits.
These checks indicate numerical stability for
the stated benchmark and discretizations.
They do not provide an interval-arithmetic
certification of the continuum integrals.

Finally, we integrate the uncoupled Majoron
background using Eq.~\eqref{eq:V_background_theta}
and evaluate the free-streaming susceptibility
from Eq.~\eqref{eq:V_chi_free}.
The local velocity response is obtained
by integrating the linearized equation
\eqref{eq:V_delta_eom}. We obtain
$\theta_0'(0)=-0.107$ and
$R_{\rm ad}=5.23\times10^{-53}$.
For the 1201-point comparison run,
$\beta(0)=4.69\times10^{-72}$ and
$Y(0)=3.73\times10^{-70}$, yielding
\begin{equation}
\begin{aligned}
|R_{\rm ret}-R_{\rm ad}|
&\lesssim3.54\times10^{-69},\\
\frac{|R_{\rm ret}-R_{\rm ad}|}{|R_{\rm ad}|}
&\lesssim6.78\times10^{-17}.
\end{aligned}
\label{eq:appE_final}
\end{equation}
The bound concerns the finite-time retarded
versus leading local current-induced scalar
response under the same prepared initial
conditions. The independent Ward-projection
correction discussed in Sec.~\ref{subsec:V_checks}
is not included in this current-difference
envelope.


\begin{thebibliography}{99}


\bibitem{Chikashige:1981ui}
Y.~Chikashige, R.~N.~Mohapatra, and R.~D.~Peccei,
``Are There Real Goldstone Bosons Associated with Broken Lepton Number?,''
Phys. Lett. B \textbf{98}, 265--268 (1981).


\bibitem{Gelmini:1980re}
G.~B.~Gelmini and M.~Roncadelli,
``Left-Handed Neutrino Mass Scale and Spontaneously Broken Lepton Number,''
Phys. Lett. B \textbf{99}, 411--415 (1981).


\bibitem{Schechter:1981cv}
J.~Schechter and J.~W.~F.~Valle,
``Neutrino Decay and Spontaneous Violation of Lepton Number,''
Phys. Rev. D \textbf{25}, 774--783 (1982).


\bibitem{Biggio:2026majoronEFT}
C.~Biggio, M.~Fuentes Zamoro, X.~Li,
L.~Merlo, and L.~Ottonello,
``How to identify a Majoron: Effective field theories
of spontaneous lepton number breaking,''
Phys. Rev. D \textbf{114}, 075023 (2026),
arXiv:2608.11522.


\bibitem{deGiorgi:2026majoron}
A.~de Giorgi, D.~Naredo-Tuero, and X.~Ponce D\'iaz,
``Majoron cosmological window: Dark matter and
thermal leptogenesis,''
Phys. Rev. D \textbf{114}, 075034 (2026).


\bibitem{Frieman:1995pm}
J.~A.~Frieman, C.~T.~Hill, A.~Stebbins, and I.~Waga,
``Cosmology with Ultralight Pseudo Nambu-Goldstone Bosons,''
Phys. Rev. Lett. \textbf{75}, 2077--2080 (1995),
arXiv:astro-ph/9505060.


\bibitem{ChunJung2024}
E. J. Chun and T. H. Jung,
``Leptogenesis driven by a Majoron,''
Phys. Rev. D \textbf{109}, 095004 (2024).


\bibitem{ChunLeeSong:2026}
E.~J.~Chun, H.~M.~Lee, and J.-H.~Song,
``Spontaneous leptogenesis in type I seesaw models,''
Phys. Rev. D \textbf{113}, 075021 (2026).


\bibitem{Simpson:2018maj}
F.~Simpson, R.~Jimenez, C.~Pena-Garay,
and L.~Verde,
``Dark Energy from the Motions of Neutrinos,''
Phys. Dark Universe \textbf{20}, 72--77 (2018),
arXiv:1607.02515 [astro-ph.CO].


\bibitem{Cheon:2026Majoron}
K.~Cheon, S.~K.~Kang, and J.~Lee,
``Majoron Dark Energy via Freezing Induced
by Quantum Coherence,''
arXiv:2607.03070 (2026).


\bibitem{Kubo:1957}
R.~Kubo,
``Statistical-Mechanical Theory of Irreversible Processes. I. General Theory and Simple Applications to Magnetic and Conduction Problems,''
J. Phys. Soc. Jpn. \textbf{12}, 570--586 (1957).


\bibitem{Berera:2008ar}
A.~Berera, I.~G.~Moss, and R.~O.~Ramos,
``Warm Inflation and Its Microphysical Basis,''
Rept. Prog. Phys. \textbf{72}, 026901 (2009),
arXiv:0808.1855 [hep-ph].


\bibitem{Lawrie2003}
I. D. Lawrie,
``Dissipation in Equations of Motion of Scalar Fields,''
Phys. Rev. D \textbf{67}, 045006 (2003).


\bibitem{Lawrie:2005friction}
I.~D.~Lawrie,
``Numerical Investigation of Friction in Inflaton
Equations of Motion,''
Phys. Rev. D \textbf{71}, 025021 (2005).


\bibitem{DAlessioPolkovnikov:2014}
L.~D'Alessio and A.~Polkovnikov,
``Emergent Newtonian dynamics and the geometric origin of mass,''
Ann. Phys. \textbf{345}, 141--165 (2014),
arXiv:1309.6354.


\bibitem{KolodrubetzEtAl:2017}
M.~Kolodrubetz, D.~Sels, P.~Mehta, and A.~Polkovnikov,
``Geometry and non-adiabatic response in quantum and classical systems,''
Phys. Rep. \textbf{697}, 1--87 (2017),
arXiv:1602.01062.


\bibitem{Enqvist:1992dm}
K.~Enqvist, K.~Kainulainen, and M.~Thomson,
``Cosmological Bounds on Dirac-Majorana Neutrinos,''
Phys. Lett. B \textbf{280}, 245--250 (1992).


\bibitem{Chang:2000pd}
D.~Chang and O.~C.~W.~Kong,
``Pseudo-Dirac Neutrinos,''
Phys. Lett. B \textbf{477}, 416--423 (2000),
arXiv:hep-ph/9912268.


\bibitem{Kobayashi:2000md}
M.~Kobayashi and C.~S.~Lim,
``Pseudo Dirac Scenario for Neutrino Oscillations,''
Phys. Rev. D \textbf{64}, 013003 (2001).


\bibitem{Biswas:2025pseudoDirac}
S.~Biswas, Vishnu~P.~K., and A.~Thapa,
``Connecting pseudo-Nambu-Goldstone dark matter
with pseudo-Dirac neutrinos in a left-right
symmetry model,''
Phys. Rev. D \textbf{111}, 075007 (2025).


\bibitem{Carloni:2026quasiDirac}
K.~Carloni, Y.~Porto, C.~A.~Arg\"uelles,
P.~S.~Bhupal Dev, and S.~Jana,
``Signatures of quasi-Dirac neutrinos in diffuse
high-energy astrophysical neutrino data,''
Phys. Rev. D \textbf{113}, 115004 (2026).


\bibitem{Dev:2023ul}
A.~Dev, G.~Krnjaic, P.~A.~N.~Machado,
and H.~Ramani,
``Constraining Feeble Neutrino Interactions
with Ultralight Dark Matter,''
Phys. Rev. D \textbf{107}, 035006 (2023),
arXiv:2205.06821 [hep-ph].


\bibitem{AdsheadCuiShelton2016}
P.~Adshead, Y.~Cui, and J.~Shelton,
``Chilly Dark Sectors and Asymmetric Reheating,''
JHEP \textbf{06}, 016 (2016).


\bibitem{AdsheadRalegankarShelton2019}
P.~Adshead, P.~Ralegankar, and J.~Shelton,
``Reheating in Two-Sector Cosmology,''
JHEP \textbf{08}, 151 (2019).


\bibitem{Babu1993}
K.~S.~Babu, I.~Z.~Rothstein, and D.~Seckel,
Planck scale symmetry breaking and Majoron physics,
Nucl. Phys. B \textbf{403}, 725 (1993),
arXiv:hep-ph/9301213.


\bibitem{GhoshLohoManna2026}
T.~Ghosh, K.~Loho, and S.~Manna,
``From $U(1)\times U(1)$ symmetry breaking to
Majoron cosmology: Insights from NANOGrav
15-year data,''
Phys. Rev. D \textbf{113}, 043036 (2026),
arXiv:2507.04342.


\bibitem{CentellesChulia2021}
S.~Centelles Chuli\'a, R.~Srivastava,
and A.~Vicente,
``The inverse seesaw family: Dirac and Majorana,''
JHEP \textbf{03}, 248 (2021),
arXiv:2011.06609.


\bibitem{Brivio2017}
I. Brivio, M. B. Gavela, L. Merlo, K. Mimasu,
J. M. No, R. del Rey, and V. Sanz,
``ALPs effective field theory and collider
signatures,''
Eur. Phys. J. C \textbf{77}, 572 (2017).


\bibitem{Bauer2021}
M. Bauer, M. Neubert, S. Renner, M. Schnubel,
and A. Thamm,
``The low-energy effective theory of axions
and ALPs,''
JHEP \textbf{04}, 063 (2021).


\bibitem{KaplanRattazzi2016}
D. E. Kaplan and R. Rattazzi,
``Large field excursions and approximate discrete
symmetries from a clockwork axion,''
Phys. Rev. D \textbf{93}, 085007 (2016).


\bibitem{GiudiceMcCullough2017}
G. F. Giudice and M. McCullough,
``A clockwork theory,''
JHEP \textbf{02}, 036 (2017).


\bibitem{GopalakrishnaTibrewala2025}
S.~Gopalakrishna and R.~Tibrewala,
``Baryogenesis from a Majorana fermion coupled to quarks,''
Phys. Rev. D \textbf{111}, 075016 (2025).


\bibitem{AkitaYamaguchi2020}
K.~Akita and M.~Yamaguchi,
``A precision calculation of relic neutrino decoupling,''
JCAP \textbf{08}, 012 (2020).


\bibitem{FrousteyPitrouVolpe2020}
J.~Froustey, C.~Pitrou, and M.~C.~Volpe,
``Neutrino decoupling including flavour oscillations and primordial nucleosynthesis,''
JCAP \textbf{12}, 015 (2020).


\bibitem{KainulainenParkkinen2024}
K.~Kainulainen and H.~Parkkinen,
``Quantum transport theory for neutrinos
with flavor and particle-antiparticle mixing,''
JHEP \textbf{02}, 217 (2024),
arXiv:2309.00881.


\bibitem{Landete:2014}
A.~Landete, J.~Navarro-Salas, and F.~Torrenti,
``Adiabatic regularization and particle creation
for spin one-half fields,''
Phys. Rev. D \textbf{89}, 044030 (2014),
arXiv:1311.4958.


\bibitem{Blaschke:2016xxt}
D.~N.~Blaschke and V.~Cirigliano,
``Neutrino Quantum Kinetic Equations: The Collision Term,''
Phys. Rev. D \textbf{94}, 033009 (2016),
arXiv:1605.09383.


\bibitem{Herranen:2009coherence}
M.~Herranen, K.~Kainulainen, and P.~M.~Rahkila,
``Towards a kinetic theory for fermions
with quantum coherence,''
Nucl. Phys. B \textbf{810}, 389--426 (2009),
arXiv:0807.1415.


\bibitem{BoyanovskyHo2004}
D.~Boyanovsky and C.~M.~Ho,
``Oscillations and evolution of a hot and dense gas
of flavor neutrinos: A quantum field theory study,''
Phys. Rev. D \textbf{69}, 125012 (2004),
arXiv:hep-ph/0403216.


\bibitem{Sigl:1993ct}
G.~Sigl and G.~Raffelt,
``General Kinetic Description of Relativistic Mixed Neutrinos,''
Nucl. Phys. B \textbf{406}, 423--451 (1993).


\bibitem{Vlasenko:2013fja}
A.~Vlasenko, G.~M.~Fuller, and V.~Cirigliano,
``Neutrino Quantum Kinetics,''
Phys. Rev. D \textbf{89}, 105004 (2014).


\bibitem{RigolinOrtizPonce2008}
G.~Rigolin, G.~Ortiz, and V.~H.~Ponce,
``Beyond the quantum adiabatic approximation:
Adiabatic perturbation theory,''
Phys. Rev. A \textbf{78}, 052508 (2008),
arXiv:0807.1363.


\bibitem{KainulainenParkkinen2024Collision}
K.~Kainulainen and H.~Parkkinen,
``Coherent collision integrals for neutrino
transport equations,''
JHEP \textbf{12}, 169 (2024),
arXiv:2407.08598.


\bibitem{Planck:2018vyg}
N.~Aghanim \textit{et al.} (Planck Collaboration),
``Planck 2018 results. VI. Cosmological parameters,''
Astron. Astrophys. \textbf{641}, A6 (2020),
arXiv:1807.06209 [astro-ph.CO].


\end{thebibliography}
\end{document}